\documentclass[traditabstract]{aa}
\usepackage{graphicx}                    
\usepackage{bm}
\usepackage{amsmath}
\usepackage{txfonts}                    
\usepackage{natbib}                     
\bibpunct{(}{)}{;}{a}{}{,}
 \usepackage{threeparttable}
 \usepackage{amssymb}                    

\def\beg{\begin{eqnarray}}
\def\ende{\end{eqnarray}}
\def\lsim{\lower.4ex\hbox{$\;\buildrel <\over{\scriptstyle\sim}\;$}}
\def\gsim{\lower.4ex\hbox{$\;\buildrel >\over{\scriptstyle\sim}\;$}}

\newcommand{\Pm}{\mbox{Pm}}
\newcommand{\Prr}{\mbox {Pr}}

\renewcommand{\vec}[1]{\mbox{\boldmath $#1$}}
\def\curl{{\rm curl}} 
\def\nuT{\nu_{\rm T}}

\def \Om  {{\it \Omega}}

\def \etaT{\eta_ {\rm T}}

\def \d{{\rm d}}
\def\gsim{\lower.4ex\hbox{$\;\buildrel >\over{\scriptstyle\sim}\;$}} 
\def\lsim{\lower.4ex\hbox{$\;\buildrel <\over{\scriptstyle\sim}\;$}} 
\def \urms{u_{\rm rms}}
\def\div{{\rm div}}

\renewcommand{\vec}[1]{\mbox{\boldmath $#1$}}

\def\aap{Astronomy \& Astrophysics}

\def\ara\&a{Ann. Rev. Astronomy Astrophysics}

\def\apjs{Astrophys. J. Suppl.}

\begin{document}



\title{Angular momentum transport by magnetoconvection and the magnetic modulation of the solar differential rotation}
\titlerunning{Angular momentum transport by rotating   magnetoconvection}
%
   \author{G.~R\"udiger\inst{1,2} \and M.~K\"uker\inst{1}}


%
  \institute{Leibniz-Institut f\"ur Astrophysik Potsdam (AIP), An der Sternwarte 16, D-14482 Potsdam, Germany,
 \and
  University of Potsdam, Institute of Physics and Astronomy, D-14476 Potsdam, Germany
}


\date{Received; accepted}
 
\abstract{
In order to explain the variance of the solar rotation law during the activity minima and maxima, the angular momentum transport by rotating magnetoconvection is simulated in a convective box  penetrated by an  inclined azimuthal  magnetic field.
Turbulence-induced kinetic  and magnetic stresses { and}  the Lorentz force of the large-scale magnetic background field are  the basic transporters of angular momentum. Without rotation, the sign of the magnetic  stresses naturally depends on the signs of the field components as  positive (negative) $B_\theta B_\phi$ transport the angular momentum poleward (equatorward). For fast enough rotation, however, the turbulence-originated Reynolds stresses start to dominate the transport of the angular momentum  flux.  The simulations show that positive ratios of the two meridional magnetic field components to the  azimuthal field 
reduce  the inward radial  as well as  the equatorward  latitudinal transport,
 which result from  hydrodynamic calculations.  
Only for $B_\theta B_\phi>0$ (generated by  solar-type rotation laws with an accelerated equator) does the magnetic-influenced  rotation at the solar surface prove to be flatter than the nonmagnetic profile  together with the observed   slight spin-down of the  equator. The latter phenomenon does not appear for antisolar rotation with polar vortex as well as for rotation laws with prevailing radial shear.
 }

\keywords{Magnetic fields  -- angular momentum transport --   convection}

\maketitle
%
\section{Introduction} \label{Section1}
The solar surface rotation law exhibits a correlation with solar activity parameters such as sunspot numbers, large-scale magnetic fields, or small bright coronal structures (SBCS). Equatorial rotation is faster during the activity minimum and slower in the activity maximum. The magnetic field seems to decelerate the solar equator  \citep{GH84,JB11}. 
This deceleration  might be interpreted as a magnetic-originated { reduction}
of the radial shear provided the radial profile $\Om=\Om(r)$ increases outwards  (`superrotation') while the lower value $\Om_{\rm in}$ is fixed by the tachocline at the bottom of the convection zone.

Simultaneously, the form of the latitudinal rotation law $\Om=\Om(\theta)$ at the surface also  varies with the phase of the activity cycle.  The latitudinal shear is reduced by the magnetic field, 
that is to say 
it is larger during the minimum activity and it is smaller during  the maximum activity \citep{RB17}.

The numbers,  however,  are small. Expressed with the traditional definition
\beg
 \Om= A + B \cos^2\theta
  \label{Om}
\ende
(with $A$ being the equatorial rotation rate and with $\theta$ being the colatitude), the variation of $A$ is about 1\%
 while $B$ varies by about 20\%  between activity minimum and activity maximum
 \citep{XS18}. The latter value implies that the  normalised equator-pole  difference of the surface rotation, $a=\delta\Om/\Om>0$, varies in time by about 3\%. We write 
 \beg
 \Delta A = A_{\rm max}-A_{\rm min}, \ \ \ \ \ \ \ \ \ \ \ \ \ \ \ \ 
 \Delta a = a_{\rm max}-a_{\rm min},
  \label{Om1}
\ende
 where the suffixes max and min concern the rotation laws with and without a magnetic field. { The observed
differences $\Delta A$ and $\Delta a$ are both negative,} hence
 \beg
 \Delta A\cdot  \Delta a>0 , \ \ \ \ \ \ \ \ \ \ \ \ \ \ \ \ \ \ \ \ \ \ \ \ \ \ 
\frac{\Delta a}{\Delta A} \simeq 3.
  \label{Om2}
\ende
The magnetic field both reduces the equatorial angular velocity and the latitudinal shear, but the magnetic-induced reduction of the equatorial velocity is weaker than that of  the pole-equator difference. We shall explain these observational findings  by calculating the  angular momentum transport in rotating convective  boxes under the influence of a strong azimuthal magnetic field combined with  weak meridional components. The Lorentz force of this field as well as  the  anisotropic Reynolds stresses  { are the main transporters of angular momentum} \citep{MP75,KR94}. We shall see that only one of the possible magnetic configurations  provides rotation laws fulfilling the conditions (\ref{Om2}). It is just this configuration ($B_\theta B_\phi>0$) which is a natural outcome of all $\alpha\Om$ dynamos which operate with  solar-type rotation laws, that is to say  with an accelerated equator. {Toroidal fields induced by negative  radial  shear produce the opposite sign, that is $B_r B_\phi<0$}. 

{With their anelastic spherical harmonic (ASH) code, \cite{BMT04} demonstrate that the simulated rotation laws at the surface of the convection zone indeed strongly vary for  magnetic models and purely hydrodynamic calculation.  Both the  pole-equator difference of the rotation rate and also its  equatorial value are reduced by the dynamo-excited magnetic fields. Also the simulations by 
{ \cite{K15}, \cite{AB15}, \cite{KKO16}, and \cite{WR18} } provide reduced equatorial rotation rates during the magnetic maximum. In some of  these calculations, the  Maxwell stress of the induced large-scale magnetic fields only plays a minor role in the transportation of angular momentum. Because of the complexity  of the numerical magnetohydrodynamic simulations (see also \cite{BM06}), it remains unclear, however, by which mechanism the magnetic field suppresses the non-uniformity of the rotation  laws which,  in  hydrodynamics, is maintained by  Reynolds stresses due to the anisotropy of the  underlying turbulence. 

\cite{BMT04} prescribe  dissipation of the angular momentum by unresolved modes by fixing a magnetically uninfluenced  eddy viscosity of $10^{12}$~cm$^2$/s, which also determines the resulting meridional flow. Consequently, the isolines of the resulting rotation law are more or less cylindrical because of the Taylor-Proudman theorem. \cite{K63} demonstrated the basic physics: If a turbulence transports angular momentum, for instance, outwards, { then the resulting super-rotating rotation law with spherical  $\Om$~isolines 
induces a clockwise meridional circulation} (in the northern hemisphere) transporting angular momentum to the equator. The   $\Om$~isolines become cylindrical as is also the case  in many 
simulations \cite[see  recent papers by][]{FM15,W18,WR18}. In mean-field models, anisotropic heat  transport due to rotating convection which produces  `warm' poles overcomes the Taylor-Proudman theorem \citep{KR95}, while in numerical simulations the poles must additionally be warmed up   \citep{BM06,MBT06}.

By a combination of local box simulations (Section 5)  and global mean-field equations (Section 6), we shall demonstrate with the present paper that the observed finding (\ref{Om2}) can be explained by the magnetic influence of the dynamo-induced large-scale fields onto the rotation-induced  non-diffusive parts of the Reynolds stresses (the `Lambda-effect').  The  Maxwell  stress of the large-scale fields (the `Malkus-Proctor effect'), however,  appears to  play only a minor role. This explanation only succeeds for positive  product $B_\theta B_\phi$ of the meridional magnetic field component $B_\theta$ and the toroidal magnetic field component $B_\phi$. { Just this condition is fulfilled if  
positive latitudinal shear on the northern hemisphere and negative latitudinal shear on the southern hemisphere is responsible for the induction of the toroidal fields,  which indeed complies with the observations. }

\section{Angular momentum transport}\label{Angular}
Both turbulence-originated Reynolds stress and  Maxwell stress must be  formulated for a turbulent fluid under the presence of a uniform background field vector $\vec B$.  The fluctuating flow   and   field components  are denoted by $\vec{u}$ and  by $\vec{b}$, respectively. The  standard Maxwell tensor
\beg
  M_{ij}= \frac{1}{\mu_0} B_iB_j - \frac{1}{2\mu_0}\vec{B}^2 \delta_{ij}
  \label{2}
\ende
of the large-scale field  turns into the generalised stress tensor
\beg
  M_{ij}^{\rm tot}= M_{ij}  +M_{ij}^{\rm T},
  \label{3}
\ende
with  the turbulence-induced  Maxwell tensor
\beg
 M_{ij}^{\rm T}= \frac{1}{\mu_0} \langle b_i(\vec{x},t)b_j(\vec{x},t)\rangle - \frac{1}{2 \mu_0}
\langle \vec{b}^2(\vec{x},t)\rangle \delta_{ij}.
  \label{3a}
\ende
The negative coefficients of the Kronecker tensors in Eq. (\ref{3}) form the total pressure which we can completely ignore  in what follows.  The difference between the
Reynolds tensor and Maxwell tensor is
\beg
 T^{\rm T}_{ij}=Q_{ij}- \frac{1}{\rho} M_{ij}^{\rm T}
  \label{3b}
\ende
with the  one-point correlation tensor being 
$
Q_{ij}= \langle u_i(\vec{x},t) u_j(\vec {x},t)\rangle
 $
of the flow. 
The expression
\beg
 T_{ij}=T^{\rm T}_{ij}- \frac{1}{\rho} M_{ij}
  \label{3c}
\ende
gives the total contribution of turbulence and magnetic background fields to the angular momentum transport. The off-diagonal  components 
\beg 
T_{r\phi}= Q_{r\phi} - \frac{1}{\mu_0\rho} \langle b_r(\vec{x},t)b_\phi(\vec{x},t)\rangle
- \frac{1}{\mu_0\rho} B_r B_\phi
\label{Tr}
\ende
 and 
 \beg
  T_{\theta\phi}= Q_{\theta\phi} - \frac{1}{\mu_0\rho} \langle b_\theta(\vec{x},t)b_\phi(\vec{x},t)\rangle
- \frac{1}{\mu_0\rho} B_\theta B_\phi
  \label{Tf}
  \ende
represent  the fluxes of specific angular momentum in the radial and latitudinal direction by the magnetoconvection under the influence of  a magnetic background field. The quantities do not contain  pressure terms. Expressions \ref{Tr} and \ref{Tf} for purely azimuthal fields have already been calculated for stellar convection under the presence of a purely azimuthal magnetic field where surprisingly a magnetic quenching of the turbulent fluxes of angular momentum did not appear \citep{KK04}.

It is known that the symmetry of the cross-correlations $ T_{r\phi}$ and $ T_{\theta\phi}$ differs from 
that  of all other components of the tensor $T_{ij}$. While $T_{r\phi}$ and $ T_{\theta\phi}$ are antisymmetric with respect to the transformation $\Om\to -\Om$, all other correlations are not. The turbulent angular momentum transport is thus odd in $\Om$, while the other terms -- the cross-correlation $T_{r \theta}$ included -- are even in $\Om$. It is easy to show that $T_{r\phi}$ is symmetric with respect to the equator if the averaged flow  and the magnetic fields are also symmetric. Then the component $T_{\theta\phi}$ is antisymmetric with respect to the equator.
These rules can only be violated if, for example, the magnetic field strengths in the two hemispheres are different.

One can easily  show that isotropic turbulence even under the influence of  rotation does not lead to finite values of  $T_{r\phi}$ and $ T_{\theta\phi}$. Only with a preferred direction $\vec g$  can a tensor $(\epsilon_{ikl} g_j+ \epsilon_{jkl}g_i)g_k\Om_l$ linear in $\Om$ exist, possessing cross-correlations with the index $\phi$ as one of the indices (if $\vec g$ is radially directed). Rigidly rotating anisotropic turbulence, therefore, is thus able to transport angular momentum (`$\Lambda$ effect'). 

As in \cite{RT86}, we write
\beg
\mu_0 M^{\rm T}_{ij}= \kappa' \vec{B}^2 \delta_{ij}-\kappa B_i B_j
\label{max1}
\ende
($\kappa>0$ for $\Om=0$) so that
\beg
\mu_0 M^{\rm tot}_{ij}= \kappa' \vec{B}^2 \delta_{ij}+(1-\kappa) B_i B_j
\label{max2}
\ende
for the sum  of small-scale and large-scale Maxwell stresses. For the nonrotating fluid, the one-point-correlation tensor may be written as
\beg
Q_{ij}=Q_{ij}^{(0)} + D' \frac{\vec{B}^2}{\mu_0\rho} \delta_{ij} + D\frac { B_i B_j}{\mu_0\rho}
\label{max1a}
\ende
with $Q_{ij}^{(0)}$ for $\vec B=0$. A vertical  magnetic field supports the vertical
turbulence intensity \citep{C61}, which is described by $D>0$ .
It follows
that\beg
T_{xz}={T}^*_{xz} + D^* \frac{B_x B_z}{\mu_0\rho}, \ \ \ \ \ \ \ \ \ \ \ \ \ 
T_{yz}={T}^*_{yz} +  D^*\frac{B_y B_z}{\mu_0\rho}
\label{max4}
\ende
with $D^*=D+\kappa-1$. In accordance with the geometry of the  box simulations presented below, here, we introduced the  coordinates $(x,y,z)$ as local Cartesian proxies of the global spherical coordinates $(r,\theta,\phi)$.

We note that the coefficients $D$ and $\kappa$ of the Reynolds stress and Maxwell stress work  in the same direction. The  sum $D+\kappa$ describes the turbulence-induced magnetic angular momentum transport which is accompanied by that of the large-scale Lorentz force, that is  to say minus 1 in the coefficient of $B_x B_z$ or  $B_y B_z$, respectively. Only the latter terms are odd in  the field components  $B_x$ and  $B_y $, but not ${T}^*$ which is even in all single field components by definition. If, therefore, two simulations exist with one and the same $B_x$, but with opposite $B_z$,  then from Eq. (\ref{max4}) it follows that ${T}^*_{xz}=(T^+_{xz}+T^-_{xz})/2$ and $D^*=\mu_0\rho(T^+_{xz}-T^-_{xz})/ 2 B_x |B_z|$. To introduce physical units for the intensity quantities, a tilde denotes   normalisation with the turbulence intensity $u^2_{\rm rms}$
calculated  without rotation and without a field. The sum $D+\kappa$
describes whether the rotating turbulence supports the angular momentum transport by the Lorentz force or not; its sign is not  fixed by definition. 

 The magnetic background field may consist of a dominating azimuthal field  $B_z$ together with weak meridional components $B_x$ and $B_y$ forming the inclination (`pitch') angles,  
\beg
p_x=\frac{B_x}{B_z}, \ \ \ \ \ \ \ \ \ \ \ \ \ \ \ \ \ \ \ \ \ \ \ \ \ \ \    p_y=\frac{B_y}{B_z},
\label{pitch}
\ende
(always taken in the northern hemisphere) with $|p_x|\ll 1$ and $|p_y|\ll 1$. We note that for the Sun,  $ p_y>0$  at the northern hemisphere  since, as generally believed,  the toroidal large-scale magnetic field $B_z$ is generated from the latitudinal  field  $B_y$ by a  latitude-dependent rotation law with an accelerated equator. Antisolar rotation leads to   $p_y<0$. The negative radial shear in the solar subsurface shear layer would produce negative values of $p_x$.  Because of the magnetic feedback, the winding-up process is almost independent of the seed field $B_y$. One  finds $B_z\simeq \delta\Om L \sqrt{\mu_0\rho}$ with $\delta\Om$ as the differential rotation  and  $L$ as the characteristic length scale so that for the Sun $B_z\lsim 10^4$~G.  Together with the observed poloidal field strength, $p_y\gsim 10^{-4}$. In the spirit of Lenz's rule, we expect that fields with positive $p_y$  lead to rotation laws with less latitudinal shear than the nonmagnetic system would   produce.
We shall indeed  show that both the inclinations $p_x$ and $p_y$ are able to reduce the nonmagnetic surface shear of the differential rotation, but only a positive $p_y$ simultaneously reduces the equatorial rotation rate as observed in the cycle maxima.

\section{No rotation}\label{No}
{
To study the effect of the magnetic field on the angular momentum transport quantitatively,
we 
ran a series of direct numerical simulations  to study convection in a rotating rectangular box. The {\sc Nirvana} code uses a Godunov scheme as described in \cite{Ziegler2004} to solve the equations of energy conservation and mass conservation, the equation of motion, and the induction equation. The model assumes a fully ionised ideal gas with constant specific heat capacity. The stratification of the gas is piecewise polytropic, \beg
\frac{\partial \ln P}{\partial \ln T}  = m+1.
\ende
A convectively unstable layer with $m=1$ was 
placed between two stably stratified layers with $m=9$. 
The heat flux is constant throughout the simulation box. The box size is $L_x \times L_y \times L_z = 2 \times 6 \times 6$ in dimensionless units. The unstable layer starts at $x=0.8$ and ends at $x=1.8$. The boundary conditions in the horizontal $y$ and $z$ directions are periodic. On the lower  boundary ($x=0$), we imposed a fixed temperature, no penetration, stress-free horizontal flow, and  pseudo-vacuum boundary conditions on the magnetic field. On the upper boundary at $x=2$, the boundary conditions imply a fixed temperature gradient, while the boundary conditions on the gas flow and the magnetic field are the same as on the lower boundary.  The mesh size is $N_x \times N_y \times N_z = 128 \times 384 \times 384.$ The simulations start with hydrostatic equilibrium and a small random perturbation and run for about 200 turnover times. 
The Rayleigh number in the unstable layer is ${\rm Ra}=10^7$. The ratio of the densities at the bottom and top of the simulation box is (only) 5.8. In code units, the isothermal sound speed is $c_{\rm ac}=100$ at the top of the unstably stratified layer, and the viscosity is $6.32\times 10^{-3}$ while those of the magnetic diffusivity and the heat diffusion coefficient are $6.32 \times 10^{-2}$. 
The magnetic field was imposed as an initial condition and periodically reset. It is constant in the vertical ($x$) direction while the boundary conditions require the horizontal field components to be zero on the upper and lower boundaries. These components therefore vary with depth, that is
\beg
B_y, B_z   \propto \sin \left ( \frac{\pi x}{2} \right).
\label{fields}
\ende
The below mentioned field amplitudes $B_y$ and $B_z$ always refer to the maxima in the centre of the box. If the 
above introduced Lorentz force factor $D$ is defined for uniform fields, then the $D$ for the simulated nonuniform fields thus slightly depends on the radial profile  used in Eq. (\ref{fields})}.
The background field was introduced as an initial condition and reset after each snapshot. This kept the vertical ($x$) and toroidal ($z$) components close to the initial configuration. For the $y$-component, however, the time between snapshots proved sufficient to develop a significant departure, which limited our ability to apply weak fields in that direction, as discussed below.
 \begin{figure}[htb]
\centering
\hskip-0.1cm
\hbox{
\includegraphics[width=4.4cm]{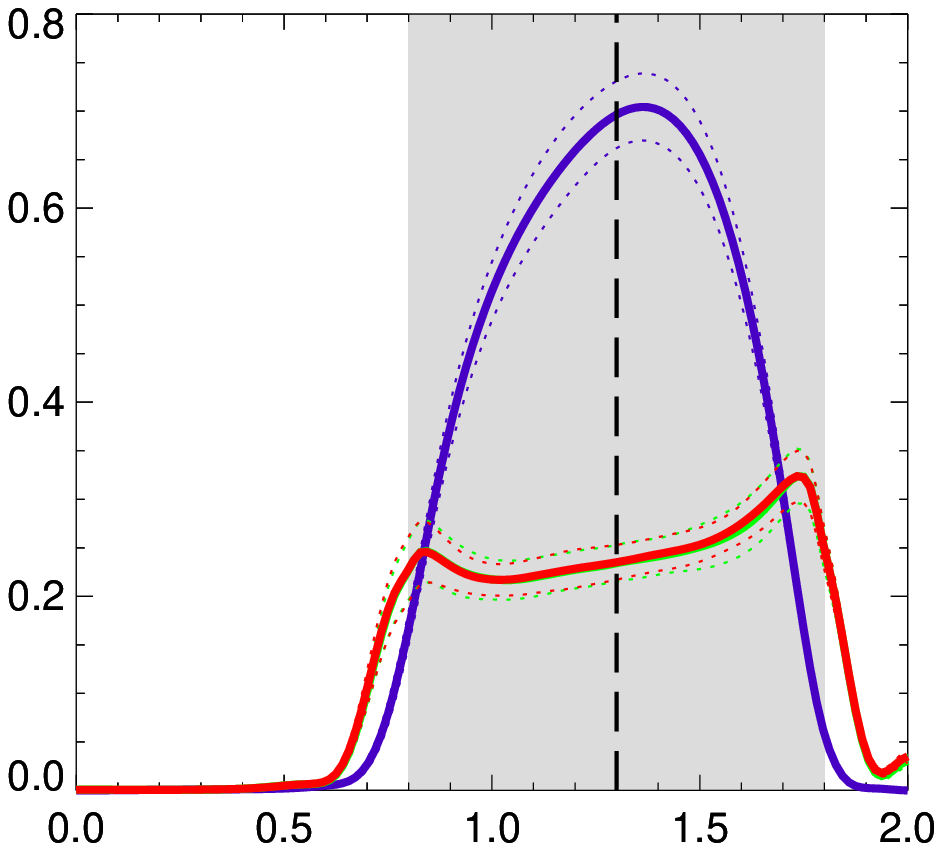}
\includegraphics[width=4.4cm]{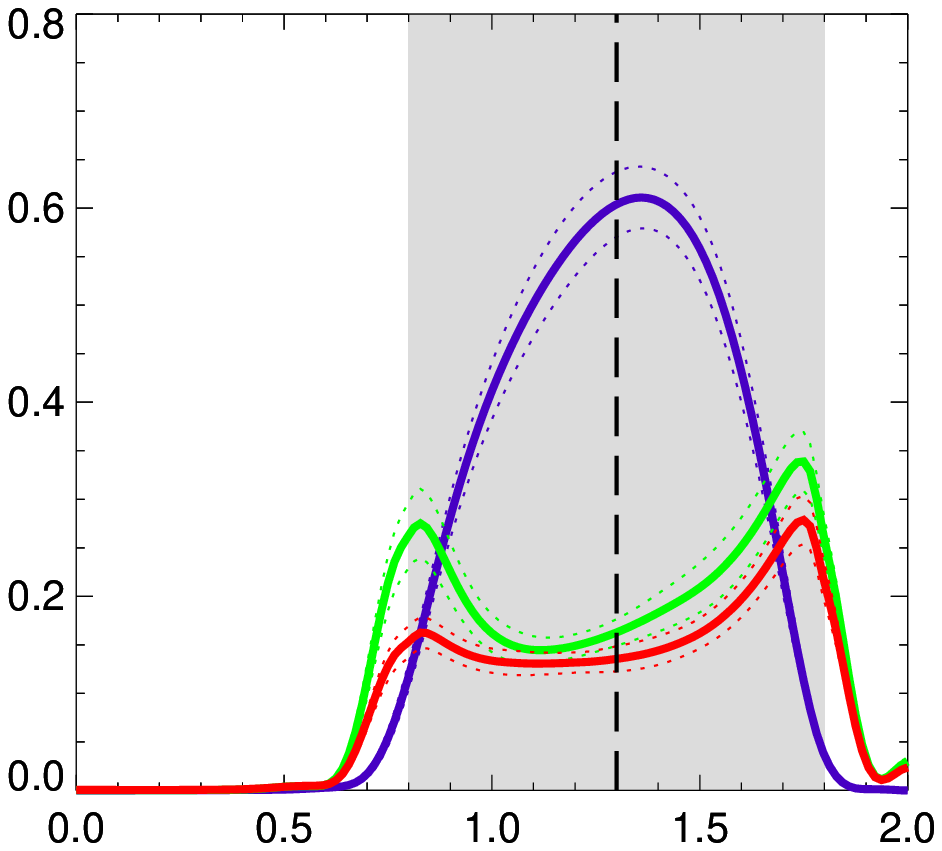}
}

\caption{
Turbulent intensities by nonrotating convection  without (left panel)  and with a magnetic field (right panel,  $\vec{B}=(0,1,10)$); $\tilde Q_{xx}$ (blue line), $\tilde Q_{yy}$ (green line), and  $\tilde Q_{zz}$ (red line) as a function of the vertical coordinate  $x$.  In the left panel, the red line hides  the green line.
The convectively unstable part of the box is grey-shaded;
$\Om=0$,  $\Prr=0.1$, and $\Pm=0.1$.}
\label{Om0}
\end{figure}
\begin{figure}[htb]
\hskip-0.3cm
\hbox{
\includegraphics[width=4.7cm]{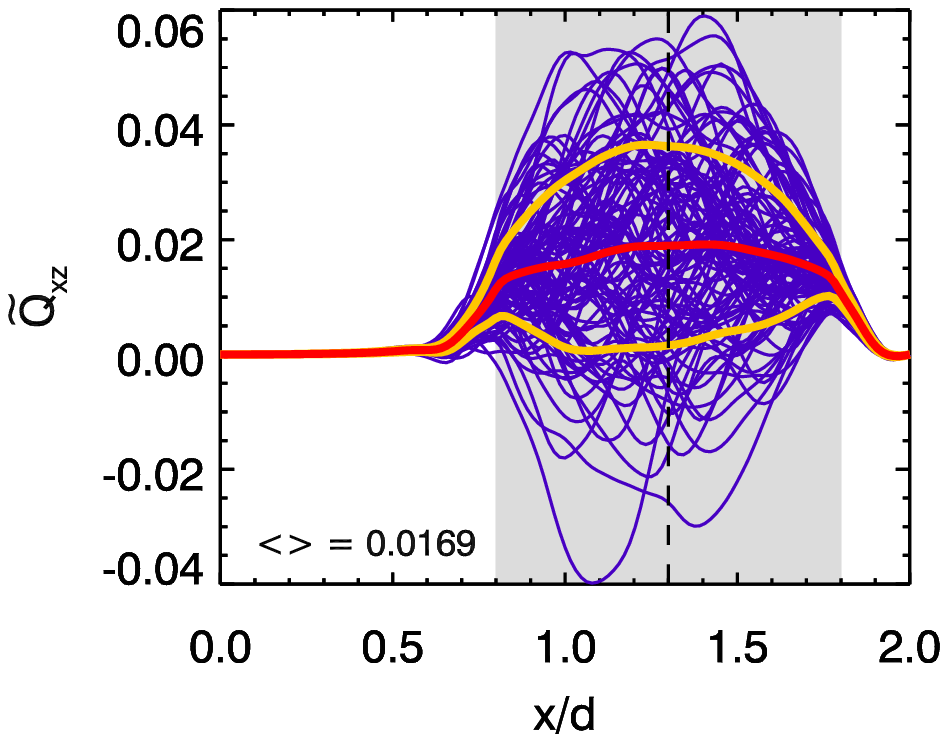}
\includegraphics[width=4.7cm]{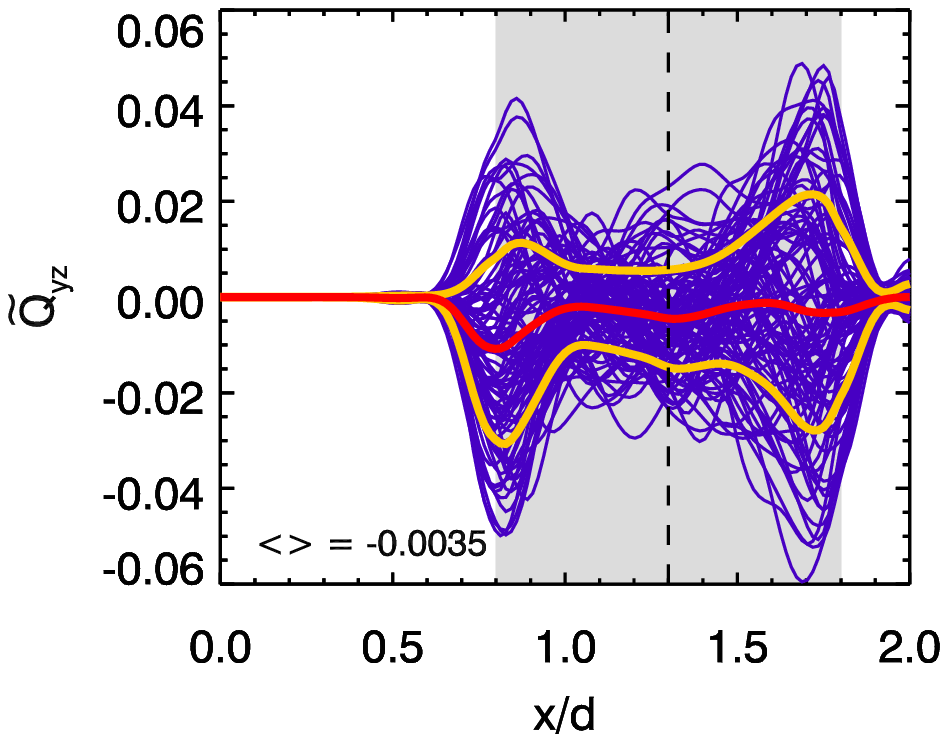}
}
\caption{Left: Vertical cross-correlation $\tilde Q_{xz}$ for  $\vec{B}=(1,0,10)$ (left panel)
and the horizontal cross-correlations $\tilde Q_{yz}$ for  $\vec{B}=(0,1,10)$ (right panel). We note that  $\tilde Q_{yz}$   is much smaller than $\tilde Q_{xz}$ (see Eq. (\ref{xz+})).
Each blue line represents the horizontal average over one snapshot. The red lines represent averages over the snapshots, and the yellow lines indicate the standard deviation. The convectively unstable part of the box is grey-shaded.  The numbers in $\langle   \rangle$-brackets are the  averages  taken over the whole unstably stratified layer and over time;
$\Om=0$,  $\Prr=0.1$, and $\Pm=0.1$.}
\label{Om0cross}
\end{figure}

Without a magnetic field,  the turbulence  in the horizontal plane is isotropic hence  the  $Q_{yy}=Q_{zz}$.  A magnetic field along a coordinate axis 
forms an anisotropy in the horizontal plane and  $Q_{yy}\neq Q_{zz}$. For the magnetic fields  $\vec{B}=(0,1,10),$ the normalised autocorrelations  $\tilde Q_{xx}, \tilde Q_{yy}$, and $\tilde Q_{zz}$ are plotted in Fig. \ref{Om0}. One finds all the intensities to be reduced by the magnetic field where the degree of quenching grows for the stronger field component (red curve). By the magnetic influence, the horizontal intensity excess $Q_{yy}-Q_{zz}$ becomes positive. The normalised numerical value is ${\tilde Q}_{yy}-{\tilde Q}_{zz}\simeq 0.045 $.

Next  the cross-correlations  $Q_{xz}$ and  $Q_{yz}$ for nonrotating magnetoconvection are discussed. As known within the quasilinear approximation, the correlation tensor of a homogeneous turbulence influenced by a weak  large-scale magnetic field $\vec{B}$ is
\beg
\begin{split}
Q_{ij}-&Q^{(0)}_{ij}=\\\ &2 \int\!\!\int  \frac{\omega^2-\nu\eta k^4 }
{(\omega^2+\nu^2 k^4)(\omega^2+\eta^2 k^4)}\frac{(\vec{k} {\vec{B}})^2}{\mu_0\rho}  \hat Q^{(0)}_{ij} \, {\rm d} \vec{k}\ 
{\rm d} \omega
\end{split}
\label{magquench}
\ende
with  $\hat Q^{(0)}_{ij}$ as the spectral tensor of the original  nonmagnetic turbulence \citep{R74}. For { isotropic} and incompressible  turbulence, the tensor is simply
\beg 
\hat Q^{(0)}_{ij}(\vec{k}, \omega)=\frac{E(k,\omega)}{16\pi k^4}
 \left(k^2 \delta_{ij}-k_i 
k_j\right), 
\label{qu} 
\ende 
where the positive-definite spectrum $E$ provides the kinetic energy
\beg 
\langle{{\vec{u}^{(0)2}}}\rangle=
\int\limits_0^\infty\!\!\int\limits_0^\infty E(k,\omega) \ 
{\rm d}k \ {\rm d}\omega. 
\label{intensity} 
\ende 
{ The insertion of Eq.\ (\ref{qu}) into Eq.\ (\ref{magquench})  and some algebra lead to}
\beg
Q_{ij}=  Q^{(0)}_{ij}- Q_3 (2 {\vec{B}}^2 \delta_{ij} -B_i B_j),
\label{iso}
\ende
where the coefficient $Q_3$ evaluates to
  \beg
Q_3= \frac{1}{60\pi\mu_0\rho}\int\!\!\int   \frac{(\nu\eta k^4-\omega^2)E(k,\omega)}
{(\omega^2+\nu^2 k^4)(\omega^2+\eta^2 k^4) }   \, {\rm d} \vec{k}\ 
{\rm d} \omega.
\label{Q3}
 \ende
One finds $Q_3>0 $ for all spectra monotonously decreasing with increasing frequency. From Eq. (\ref{iso})  for the magnetic-induced anisotropies in the horizontal plane follows
\beg  
Q_{yy}-Q_{zz} = Q_3 (B_y^2- B_z^2).
\label{xx}
\ende
The turbulence intensity should be  increased in the direction of the magnetic field. For dominating $B_z$, the difference proves to  be negative, but it is positive in our simulations (see Fig. \ref{Om0}). 
For the cross-correlation, one obtains the simple relations
   \beg
Q_{xz} = Q_3 B_x B_z, \ \ \ \ \ \ \ \ \ \ \ \ \ \ \ \ \ \ \ \ \ \ \ \ \  Q_{yz} = Q_3 B_y B_z.
\label{xz}
 \ende
For positive field components, both cross-correlations should be positive, but only the first of these relations is confirmed by the numerical simulations shown in Fig. \ref{Om0cross}. 

The reason for the discrepancies is the anisotropy of the original turbulence field.  For anisotropic turbulence fields, the above model is too simple. To the spectral tensor as shown in Eq. (\ref{qu}), an anisotropic  turbulence  may be  added which, for example,  has no velocity components in the  $\vec g$ direction, that is 
\beg 
\begin{split}
&\hat Q^{(0)}_{ij}(\vec{k}, \omega)=
\frac{E_\perp(k,\omega)}{16\pi k^4}\times\\\ &
 \left((k^2-(\vec{g}\vec{k})^2) (\delta_{ij}-g_i g_j)   -(k_i -(\vec{gk})g_i)(k_j -(\vec{gk})g_j)\right). 
\end{split}
\label{quh} 
\ende 
A turbulence with a dominating intensity in the $\vec g$-direction requires  $E_\perp<0$, while  $E_\perp>0$ originates a horizontal turbulence. Furthermore, $E_\perp< 0 $ is a necessary condition to describe anisotropic  turbulence with dominating vertical intensities. 

The total correlation tensor, for example as can be seen in Eq. (\ref{iso}), is completed by
\beg
\begin{split}
Q_{ij}=  ... - Q_2 (\frac{3}{2} &{\vec{B}}^2(\delta_{ij}-g_ig_j)
-(\vec{g}\vec{B})^2 \delta_{ij} \\\ &+(\vec{g}\vec{B})(g_i B_j+g_j B_i) -B_i B_j)
\end{split}
\label{aniso}
\ende
with $Q_2$ similar to (\ref{Q3}) but with $E_\perp$ instead of $E$. Furthermore, $Q_2$ is not positive-definite. We note that the influence of the magnetic field does not provide vertical motions if the turbulence is strictly horizontal. 
With this turbulence model, Eq. (\ref{xx}) becomes
\beg
Q_{yy}-Q_{zz} =-(Q_3+Q_2) B_z^2. 
\label{xxzz}
\ende
Here again $|B_z|$ is assumed to be much larger than the other components. Obviously,  the magnetic field in the $z$ direction can increase the anisotropy of the turbulence in favour of the  $z$-component of the velocity. The effect vanishes, however, for $Q_2+Q_3\lsim  0$, that is for turbulence with a dominant vertical intensity. 
After Eq. (\ref{xxzz}), the magnetic-induced anisotropy of the turbulence intensity in the horizontal plane changes its sign with  the sign of  $Q_2+Q_3$. 

 
For the cross-correlations, that is Eq. (\ref{xz}), it now follows that
\beg
Q_{xz} = Q_3 B_x B_z, \ \ \ \ \ \ \ \ \ \ \ \ \ \ \ \ \ Q_{yz} = (Q_2+Q_3) B_y B_z,
\label{xz+}
 \ende
where  the $x$-direction is parallel to  $\vec g$.  The first expression remains uninfluenced by the anisotropy, while the second one no longer has a definite sign. We note that the cross-correlations $Q_{xz}$ and  $Q_{yz}$ in anisotropic turbulence differ  even for identical field strengths and pitch angles. The vertical cross-correlation  $Q_{xz}$ exceeds the horizontal cross correlation  $Q_{yz}$ if the vertical   turbulence intensity exceeds the horizontal  one and  vice versa.
 For turbulence with a dominating intensity in the $\vec g$-direction, it is $Q_3+Q_2\lsim  0$ that leads to $Q_{yz}\lsim 0$, which is
 confirmed     by the simulation results of  Fig. \ref{Om0cross} with its  rather  small numerical values of the horizontal cross-correlation $Q_{yz}$.
The correlation $Q_{xz}$ is only due to the isotropic part of the turbulence field and should never vanish if $B_x$ and $B_z$ do not vanish. It only vanishes in a coordinate system
with one of the axes parallel to  the total magnetic field vector.

The ratio of the cross-correlation to the excess of the autocorrelations results from Eqs. (\ref{xxzz}) and (\ref{xz+}) simply to
\beg
\frac{Q_{yz}}{Q_{yy}-Q_{zz}} = - \frac{B_y}{B_z},
\label{rat}
 \ende
 which after the numerical results  in the right panels of Figs. \ref{Om0}  and \ref{Om0cross}  is well fulfilled. The numbers may also demonstrate the high accuracy of our numerical code. 
 
As the next step,  the total horizontal cross  component $T_{yz}$  may be  calculated for the magneto-convective fluid with no rotation but with $p_y=\pm 0.1$.
The total stress can be written as $T_{yz}=Q_{yz}-m_{yz}-M_{yz}$ with $Q$ as  the Reynolds stress, $m$ as the small-scale Maxwell stress, and $M$ as the large-scale Lorentz force $M_{yz}=B_yB_z/\mu_0\rho$.
\begin{figure}[htb]
\hskip-0.3cm
\vbox{
\hbox{
\includegraphics[width=4.7cm]{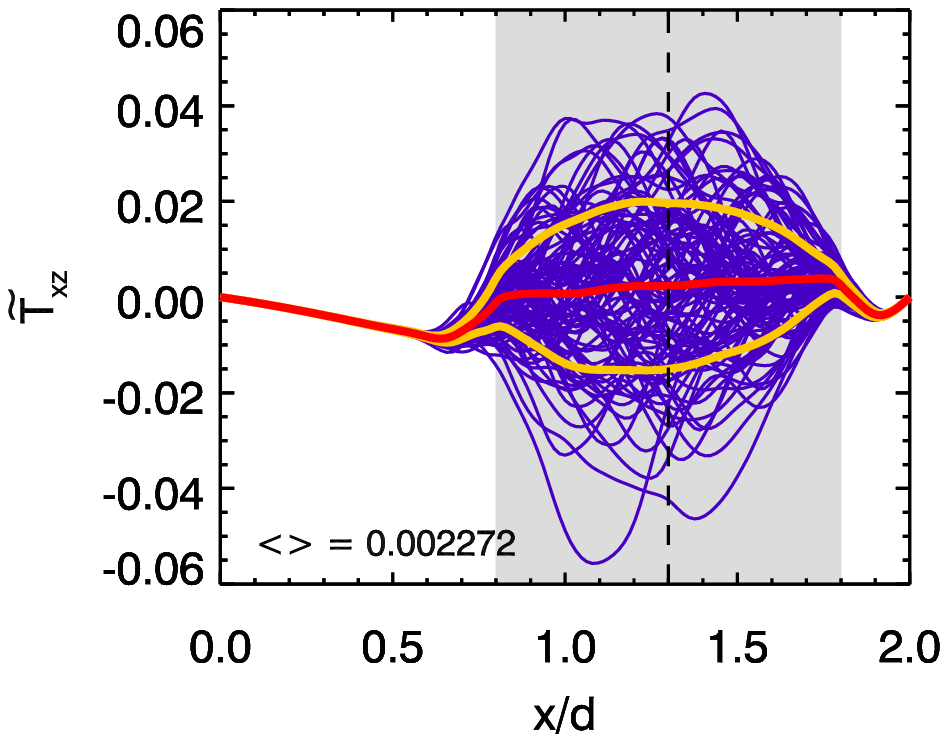}
\includegraphics[width=4.7cm]{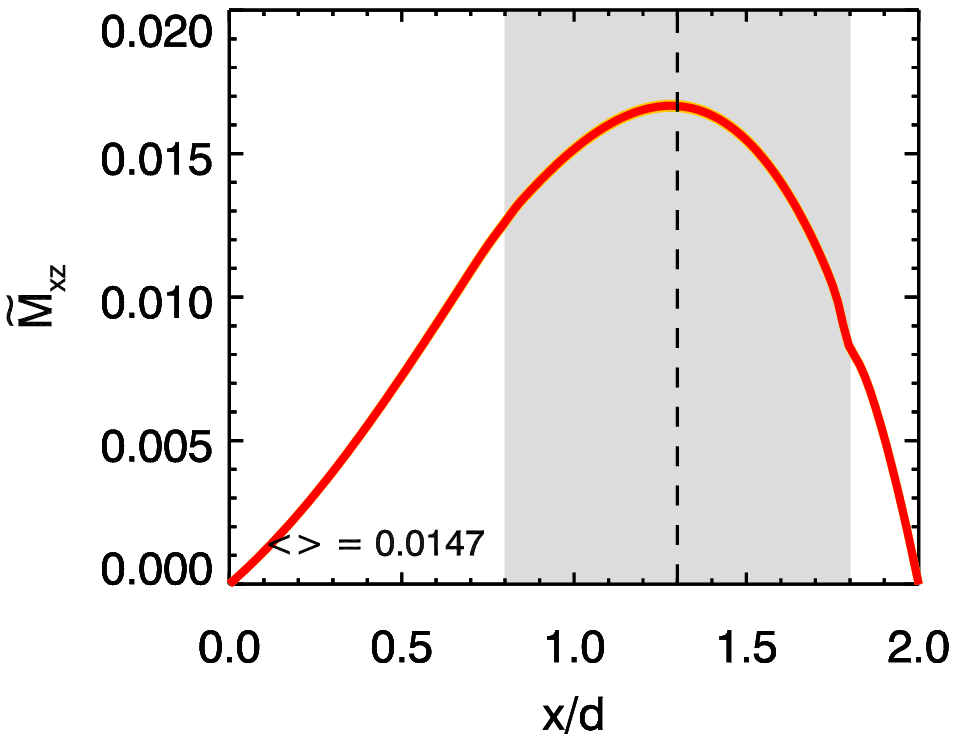}
}
\hbox{
\includegraphics[width=4.7cm]{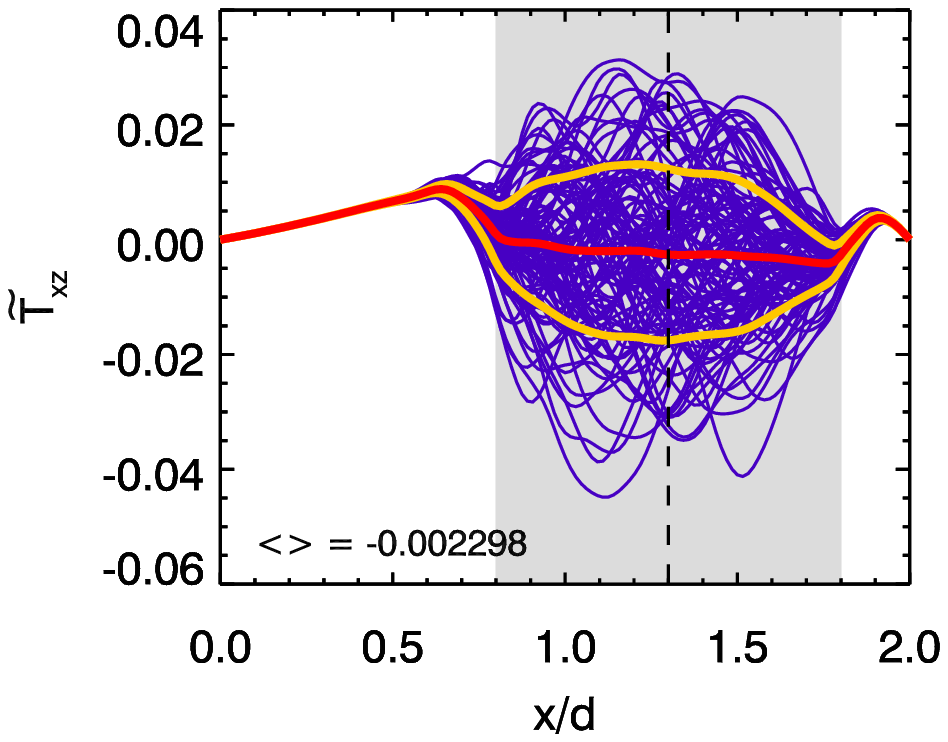}
\includegraphics[width=4.7cm]{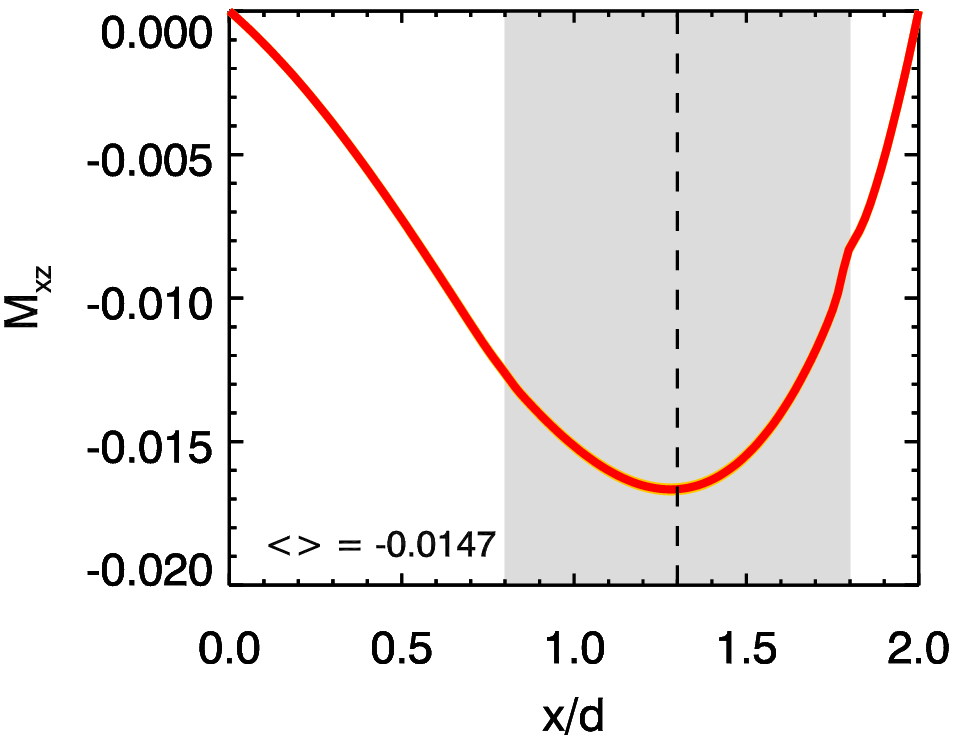}
}
}
\caption{Snapshots of the $\tilde T_{xz}$-correlations (left) and the large-scale Maxwell stress $\tilde M_{xz}$ (right) -- normalised with $u_{\rm rms}^2=28$ -- for a nonrotating fluid with  magnetic fields 
 $\vec{B}=(1,0,10)$ (top) and  $\vec{B}=(1,0,-10)$ (bottom).  We find the resulting torque $\tilde T_{xz}$ to be rather small since Reynolds stress and Maxwell stress nearly cancel each other.  The signs of the quantities in the top and the bottom  are different; $\Om=0$,  $\Prr=0.1$, and $\Pm=0.1$. }
\label{Om0vert}
\end{figure}
\begin{figure}[htb]
 \hskip-0.3cm
\vbox{
 \hbox{
\includegraphics[width=4.7cm]{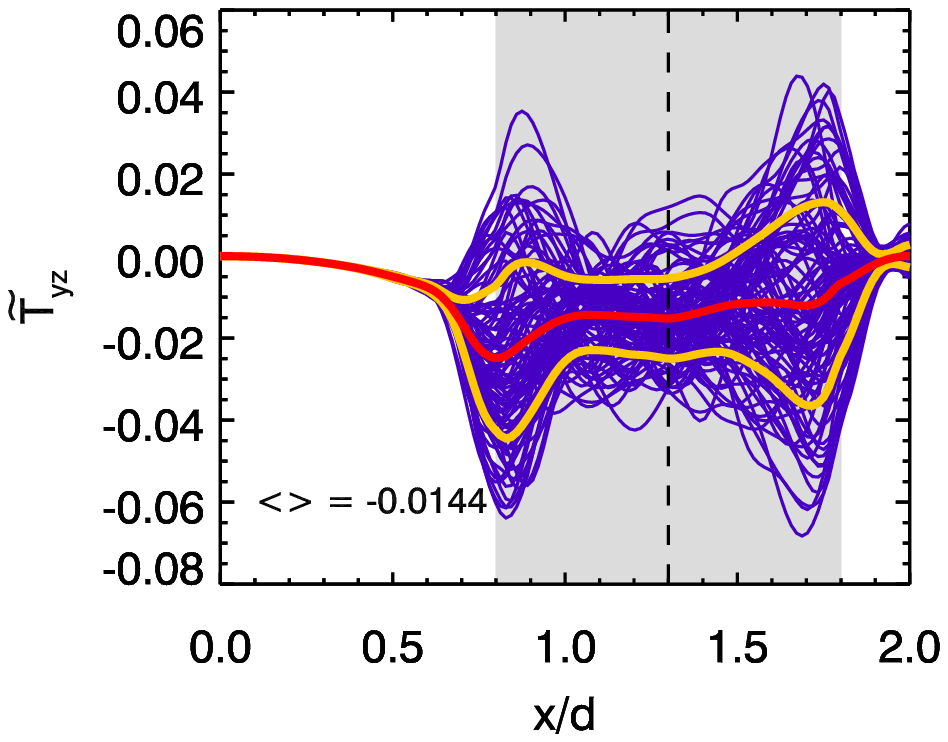}
\includegraphics[width=4.7cm]{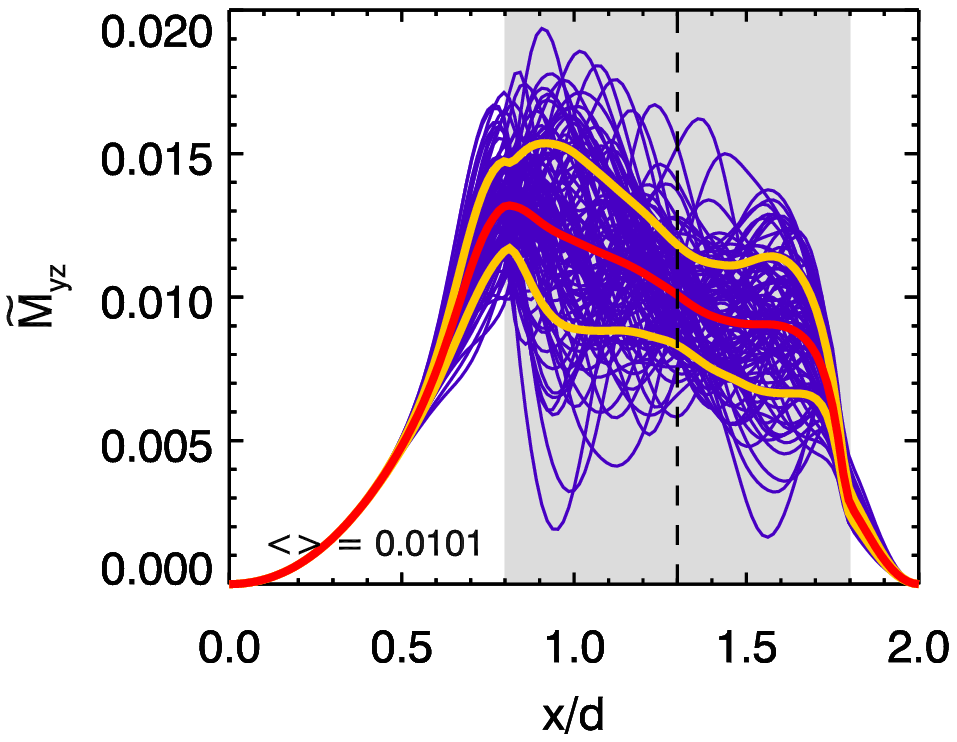}}
\hbox{
\includegraphics[width=4.7cm]{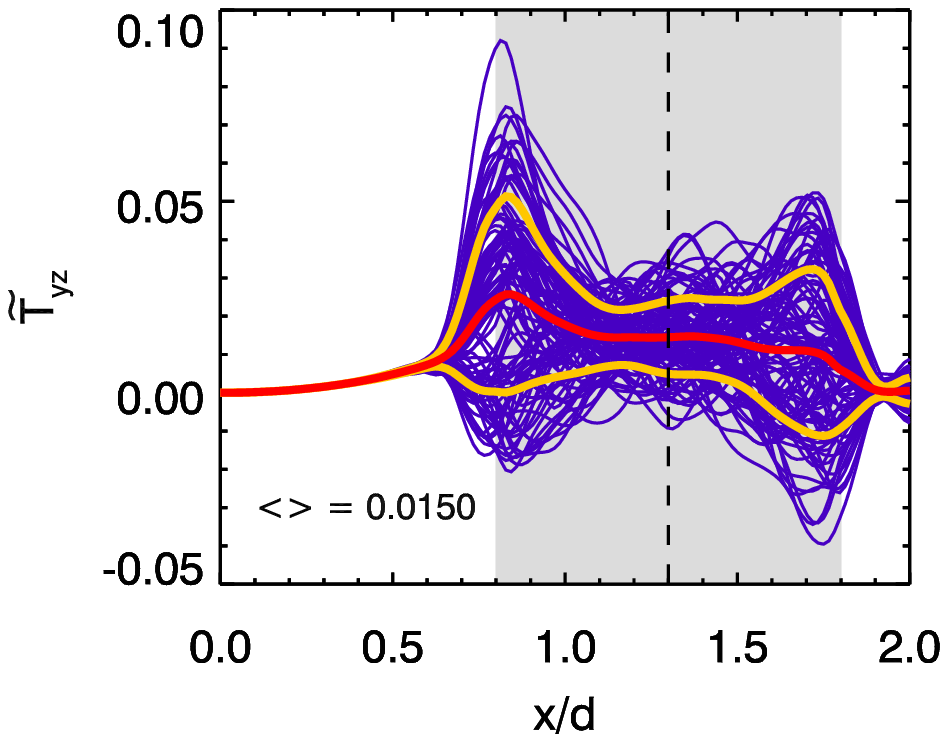}
\includegraphics[width=4.7cm]{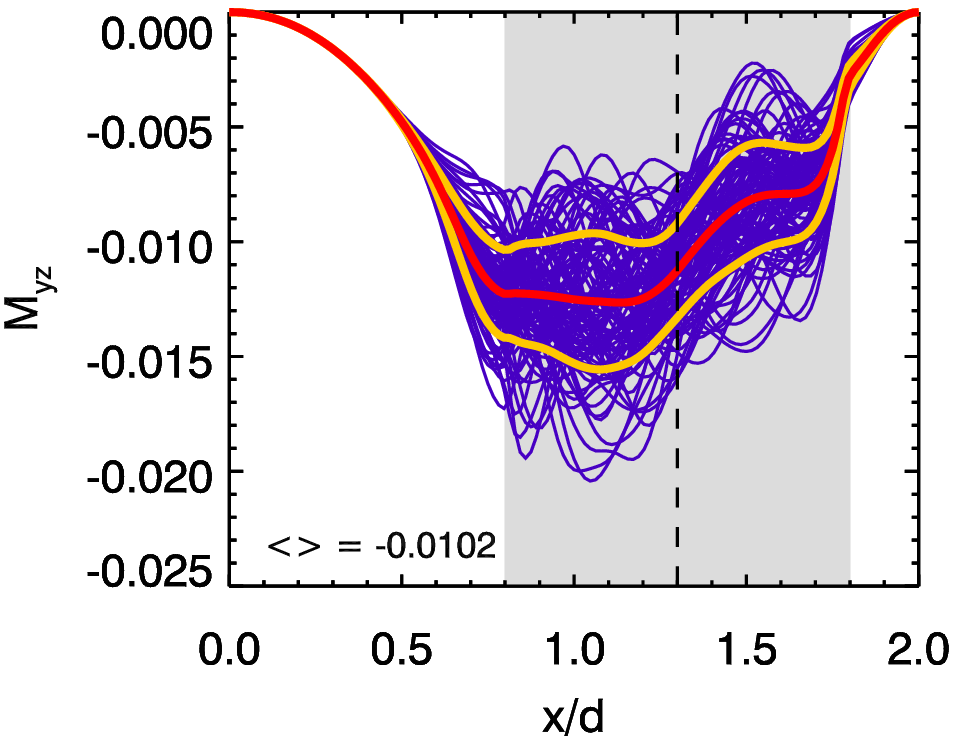}}
}
\caption{Similar to Fig. \ref {Om0vert}, but for the horizontal torque $\tilde T_{yz}$ for a nonrotating fluid with  a magnetic field $\vec{B}=(0,1,10)$ (top) and  $\vec{B}=(0,1,-10)$ (bottom). 
It is $\tilde T_{yz}\simeq -\tilde M_{yz}$ hence the Reynolds stress is not important. 
Again the signs of the quantities in the top and the bottom  are different;
$\Om=0$,  $\Prr=0.1$, and $\Pm=0.1$.}
\label{Om0hor}
\end{figure} 

As shown by Figs. \ref{Om0vert}  and  \ref{Om0hor}, the  off-diagonal elements $T_{xz}$ and $T_{yz}$  due to  the Reynolds stress and the  Maxwell stress exist, but their typical behaviour is different. In contrast to Eq. (\ref{xz}) for isotropic turbulence,  the two  cross-correlations are not equal by far;   $T_{xz}$ is  much smaller than   $T_{yz}$. The reason is that $Q_{xz}\simeq M_{xz}$ so that $T_{xz}\simeq 0$,  but this is not true for the
 horizontal  correlation $(yz)$. In this case, the Reynolds stress is much smaller than  the Maxwell stress hence 
 $T_{yz}\simeq - 1.5 M_{yz}$.  The horizontal angular momentum transport is mainly  due to the large-scale Lorentz force. The contribution of the Reynolds stress to the total stress is only 25\% and that of the turbulent  Maxwell stress is only 5\%. Because of the averaging procedure, the numerical result $\tilde M_{yz}\simeq 0.010$ for $B_y=1$ and $B_z=10$ is smaller than $\tilde M_{yz}\simeq 0.028$  obtained with the amplitudes of $B_y$ and $B_z$. For simplicity in the following, we shall use the latter expression so that  the applied magnetic fields are slightly overestimated compared with the  averaged fields.

For   toroidal fields  $B_z$ with  different  signs,  the two fluxes   $\tilde T_{yz}$ and  $\tilde T_{yz}$ also differ by the sign (Fig. \ref{Om0hor}). Without rotation, we find $\tilde D^*\simeq - 0.018$, that is $D^*\simeq -0.5 $  and $D+\kappa\simeq 0.5$. 
The density stratification reduces $D^*$ by 50 \%, that is  from -1 to -0.5. For rotating boxes, the question remains as to how much the magnetic-influenced   Reynolds stress influences these results.


\section{Helicity and alpha effect}
In {rotating} boxes also finite values of the $\alpha$ effect should evolve with consequences for the effective large-scale fields. The $\alpha$ effect generates poloidal field components   $B_y$ from the toroidal field component $B_z$, which is always the largest one in the simulations. If, as in our models, only the radial coordinate $x$ serves for the calculations of gradients, then the only relevant  component of the induction equation is
\beg
\frac{\partial B_y}{\partial t}= -\frac{\d}{\d x}(\alpha B_z) + \etaT \frac{\d^2 B_y}{\d x^2},
\label{ind11}
\ende
where the $\alpha$ represents the ($\phi\phi$) component, which is the most important one for the $\alpha\Om$ dynamo mechanism, and the $\etaT$ is the eddy diffusivity. We know that intensity stratifications provide finite values of the $\alpha$ effect  as  in the well-known relation  $\alpha_{\phi\phi}= -\hat\alpha {\vec  \Om}\cdot\nabla u_{\rm rms}$ with positive $\hat\alpha$. Because of the radial boundary conditions, the gradient of $\urms$ must be negative (positive) at the top (bottom) of the box. In the northern hemisphere, the resulting $\alpha$ effect is thus  always positive (negative) at the top (bottom) of the box. The same argument holds for the helicity $\cal H= \langle \vec{u}\cdot\curl~\vec{u} \rangle$, but with opposite signs: negative at the top and positive at the bottom. {  We computed the helicity for fast rotation 
($\Om=10$)  of the magnetised fluid.  As expected, the quantity $\cal H$ is positive (negative) in the lower (upper) part of the numerical box. It vanishes almost exactly at its centre. }  
 We also note,   as expected, that the helicity at the poles { exceeds} the helicity of mid-latitudes (Fig. \ref{heli1}). 
\begin{figure}[h]
\hskip-0.5cm
 \hbox{
  \includegraphics[width=4.7cm]{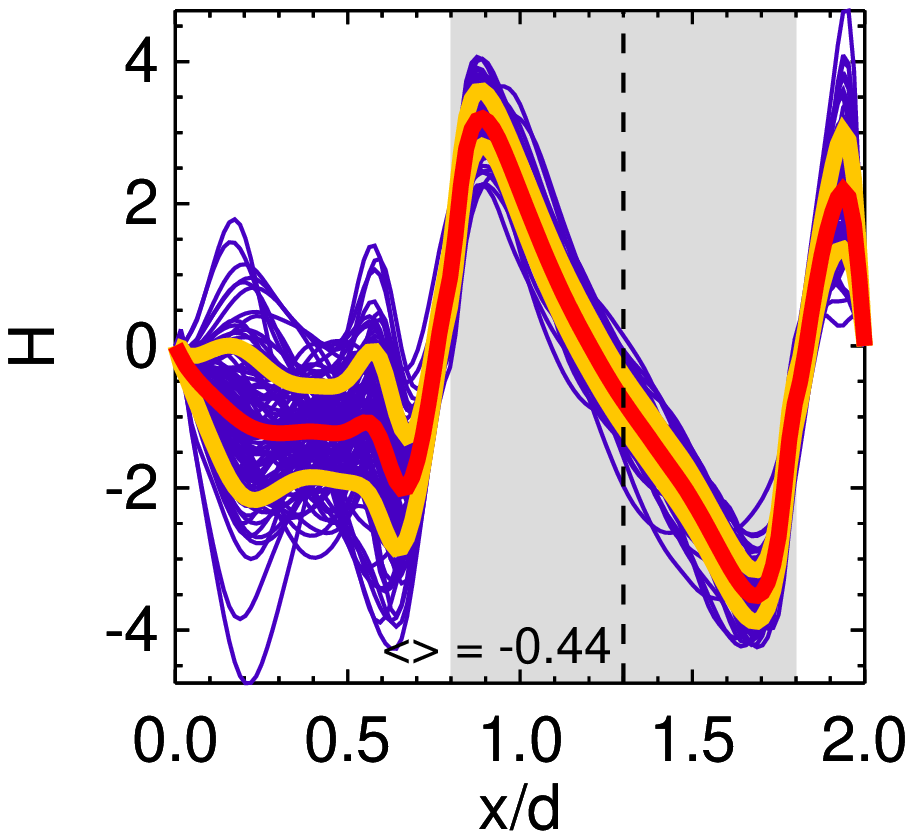}
 \includegraphics[width=4.7cm]{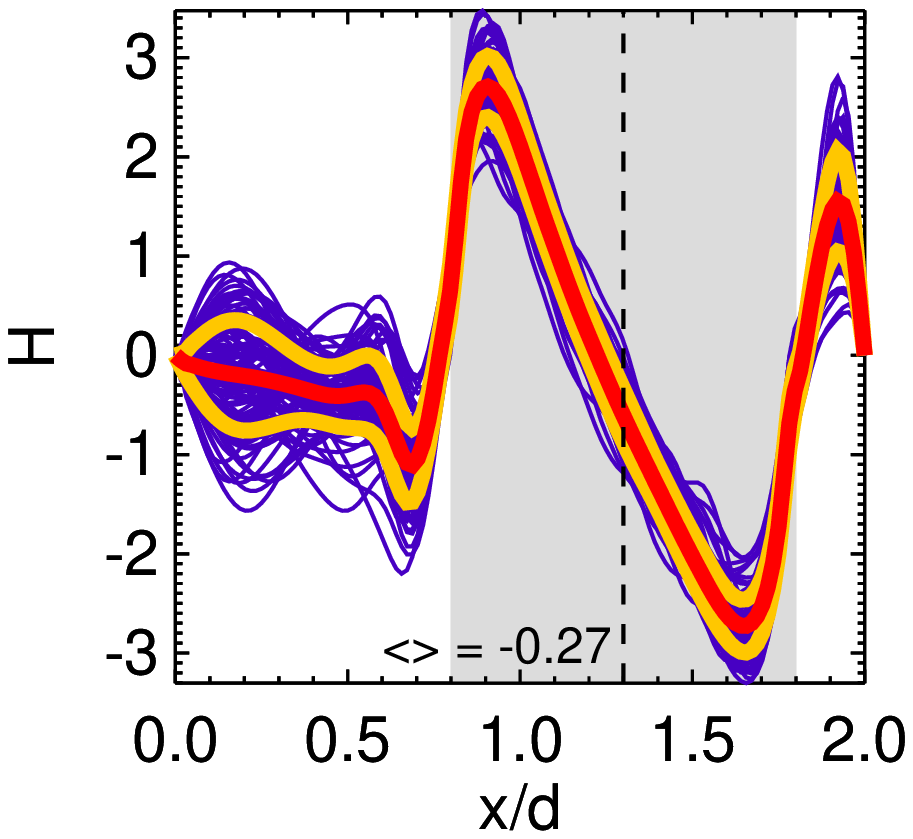}}
\caption{Snapshots of the small-scale helicity $\cal H$ at the north pole ($\theta=0^\circ$, left) and mid-latitudes ($\theta=45^\circ$, right) for fast rotation. As expected, the helicity vanishes in the middle of the box. The applied magnetic field is $\vec{B}=(0,0.1,10)$,
$\Om=10$,  $\Prr=0.1$, and $\Pm=0.1$.}
\label{heli1}
\end{figure} 

The rotation rate   $\Om\simeq 10$ represents a Coriolis number of $\Om^*= 2 \tau_{\rm corr} \Om\simeq 2$  if the correlation time $\tau_{\rm corr}\lsim 0.1$ is used, which resulted from an autocorrelation analysis of the same numerical convection  model \citep{KR18}. We note that $\Om^*\simeq 2$ is close to the solar value. With the same correlation time, the helicity amplitude of 2.5 in Fig. \ref{heli1} leads to  $\alpha/u_{\rm rms}\simeq 0.05$, which is slightly  smaller than what 
\cite{OS01,OS02}  found with 10\%  for rotating magnetoconvection.
 In their simulations,  \cite{KK09} also found typical values of the order of 10\% (see their Fig. 8).

One thus finds  unavoidable consequences for the simulations. The $\alpha$ effect automatically generates  meridional  large-scale fields $B_y$ from the toroidal field $B_z$ in accordance with the stationary induction equation (\ref{ind11}), that is $\etaT\d^2 B_y/\d x^2=B_z \d\alpha/\d x $. For positive $B_z$ and positive $ \d\alpha/\d x$ ($\alpha$ growing with height), the second derivative of $B_y$ is obviously  positive. The small-scale $\alpha$ effect, therefore, generates  a minimum of $B_y$ between the top and bottom of the box.  
The Maxwell stress   $B_y B_z$, therefore,  is  reduced compared to its value without rotation; it may even change its sign.  The right panel of Fig. \ref{heli2} demonstrates how the applied value of $B_y=0.10$ is changed to $B_y=-0.16$ by the action of the $\alpha$ effect  so that the simulated cross-correlation $T_{yz}$ is larger than the `real' one without an $\alpha$ effect.  

The right panel of Fig. \ref{heli2}  shows negative $B_y$ in the bulk of the convective box despite the small and positive starting field. Test calculations with $B_z=-10$  provided  the opposite sign of the $B_y$.  On the other hand, the radial component $B_x$ proves to be completely uninfluenced  (left panel of Fig. \ref{heli2}).  The   stress components $T_{xz}$ are thus not  modified by  the  small-scale $\alpha$ effect resulting from rotating fluids.
\begin{figure}[hbt]
 \centering
 \hskip+1cm
 \hbox{
 \includegraphics[width=4.5cm]{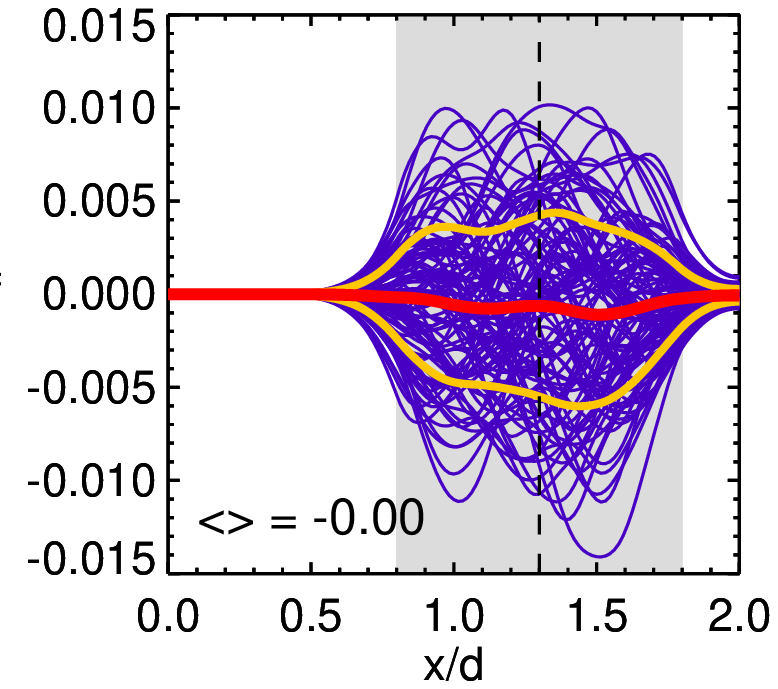}
\includegraphics[width=4.5cm]{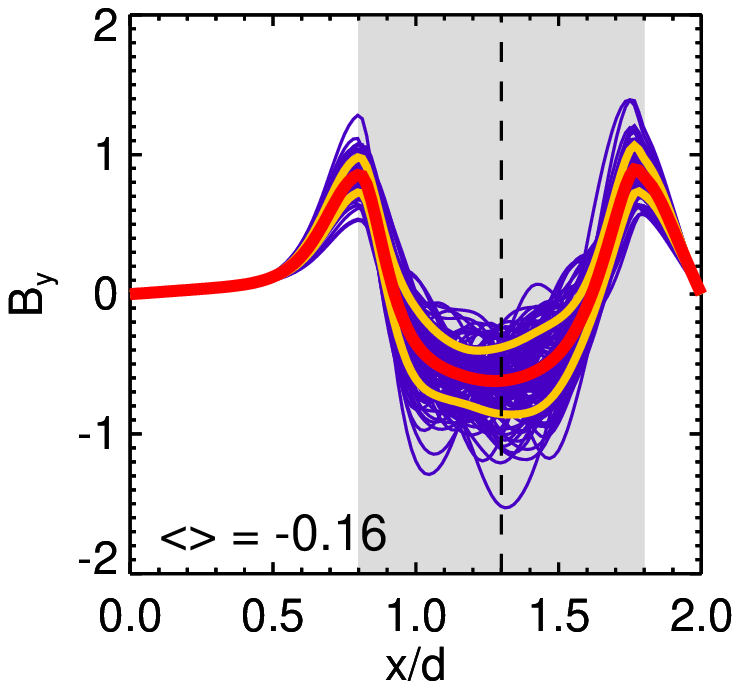}}
\caption{Snapshots of the large-scale field components $B_x$ (left) and $B_y$ (right) 
in the box. We note the transformation of $B_y>0$ to $B_y<0$ by the inbox $\alpha$ effect. The radial field $B_x$ is not concerned; $\vec{B}=(0,0.1,10)$, $\Om=10$, 
  $\Prr=0.1$, $\Pm=0.1$, and $\theta=45^\circ$.}
\label{heli2}
\end{figure}

To sum up, by   the $\alpha$ effect,   the latitudinal angular momentum transport is formally increased  for the positive inclination angle of the magnetic field (in the northern hemisphere).
The `real' stress values without this  impact are thus smaller than the simulated values. The  $T_{yz}$ given  in Table \ref{tab2} (and also in Table  \ref{tab1}) are thus maximal values. The corrections, however, are only small as for $p_y\ll 1$ and $p_y\gg1 $; the rotation-induced terms are basically larger than the Maxwell stress $B_yB_z$  for $\Om=0$. They are in particular small for models with $|p_y|\gg 1$.

{ In Table \ref{tab2}  data are given leading to the latitudinal flux  $T_{yz}$ of angular momentum.
Different values of $T_{yz}$ numerically result for  the fields ${\vec B}=(0, 0.1, 20) $ and ${\vec B}=(0, 20, 0.1) $ if the box rotates with $\Om=10$. In all cases,  $|B_yB_z|=2$. The magnetic fields of all  examples have the same total value of about 20. For the models in the upper two lines,  the pitch angle $p_y$ is positive. Also for negative $p_y$, that is  for the fields ${\vec B}=(0, 0.1,-20) $ and ${\vec B}=(0, 20,-0.1)$ in the lower two lines,  the results differ. The fluxes  are only  equal for  zero rotation  (the last column of the table). The  differences are  thus due to the $\alpha$ effect. 
They also  differ for positive  and for negative $p_y$, as it should; the flux is increased for negative $B_y B_z$.} 

It should
be mentioned that  the magnetic amplitude $B_z=20$
represents a field  almost in  equipartition with the calculated  turbulence rms velocity   $u_{\rm rms}\simeq 5.3$. We also conclude from    the numbers in Table \ref{tab2} that for saturated magnetic fields  the angular momentum transport by  the Lorentz force is by far overcompensated for by the  angular momentum transport of the rotating turbulence. The total stress is positive for all the given   magnetic field examples, but it is smallest for the field with the positive solar value $p_y=5\cdot 10^{-3}$.  The Lorentz force numbers for $\Om=0$ given in the last column of Table \ref{tab2} only contribute to about  10\% of the total stresses.   For positive $p_y$, the $T$ values 
with and without rotation even possess opposite signs. It becomes obvious here that the Malkus-Proctor effect does not play a major role in the angular momentum transport in stellar convection zones.
 \begin{table}
\caption{
 Latitudinal angular momentum transport $\tilde T_{yz}$ for very large or very small inclination angles,  $p_y>0$ in the upper two lines and $p_y<0$ in the lower two lines. The last column gives the contribution for $\Om=0$, i.e., without both the $\alpha$  and $\Lambda$ effect; $\Om=10$, $\Prr=\Pm=0.1$, and $\theta=45^\circ$.
}
\label{tab2}
\centering
\begin{tabular}{lcc|cc}
\hline\hline

\\

$B_z$ & $B_y$ &$p_y$ &$\tilde T_{yz}$ &   $\tilde T_{yz}|_{\Om= 0}$\\ \\
\hline
&&&&\\
20 & 0.1 &$5\cdot 10^{-3}$& 0.0159 &   \\
&&& &- 0.0030\\
0.1 &20 &200& 0.0285  & \\ \\
 -20& 0.1 &$-5\cdot 10^{-3}$ & 0.0255  & \\
 &&&&0.0030\\
 -0.1 & 20 &-200& 0.0334  & \\
  &&&&\\
  \hline
\end{tabular}
\end{table}
\section{Rotating magnetoconvection}
In a stratified medium, the (solid-body)  rotation alone provides finite  cross components of the Reynolds stress without any magnetic fields. 
The magnetic field, however,  modifies these correlations and also enhances or reduces the angular momentum transport by the generation of  Maxwell stresses. 

 Without magnetic fields, a rotation with $\Om=10$  produces  cross-correlations of the normalised values $\tilde Q_ {xz}= -0.062$ and  $\tilde Q_ {yz}= 0.024$ (from Table \ref{tab1}, first line) at a latitude
of $\theta=45^\circ$. With numbers taken from Fig. \ref{Om0}  in both cases, the correlation coefficient $c=|Q_{ij}|/\sqrt{u_i^2 u_j^2}$ is about 0.5 for the given cross-correlations (see Fig.\ref{Om0}).  For fast rotation, the hydrodynamic radial $\Lambda$ effect is  negative (the angular momentum flows inwards), while the latitudinal   $\Lambda$ effect is always positive (the angular momentum flows equatorward). Basically, it is $|Q_{xz}|\gsim Q_{yz}$. The numerical values strongly depend on the latitude. By definition, both  correlations vanish at the poles and    $T_{yz}$  also vanishes at the equator (Fig. \ref{lambda}). 
\begin{figure}[hbt]
  \centering
 \hskip-0.2cm
 \hbox{
 \includegraphics[width=4.5cm]{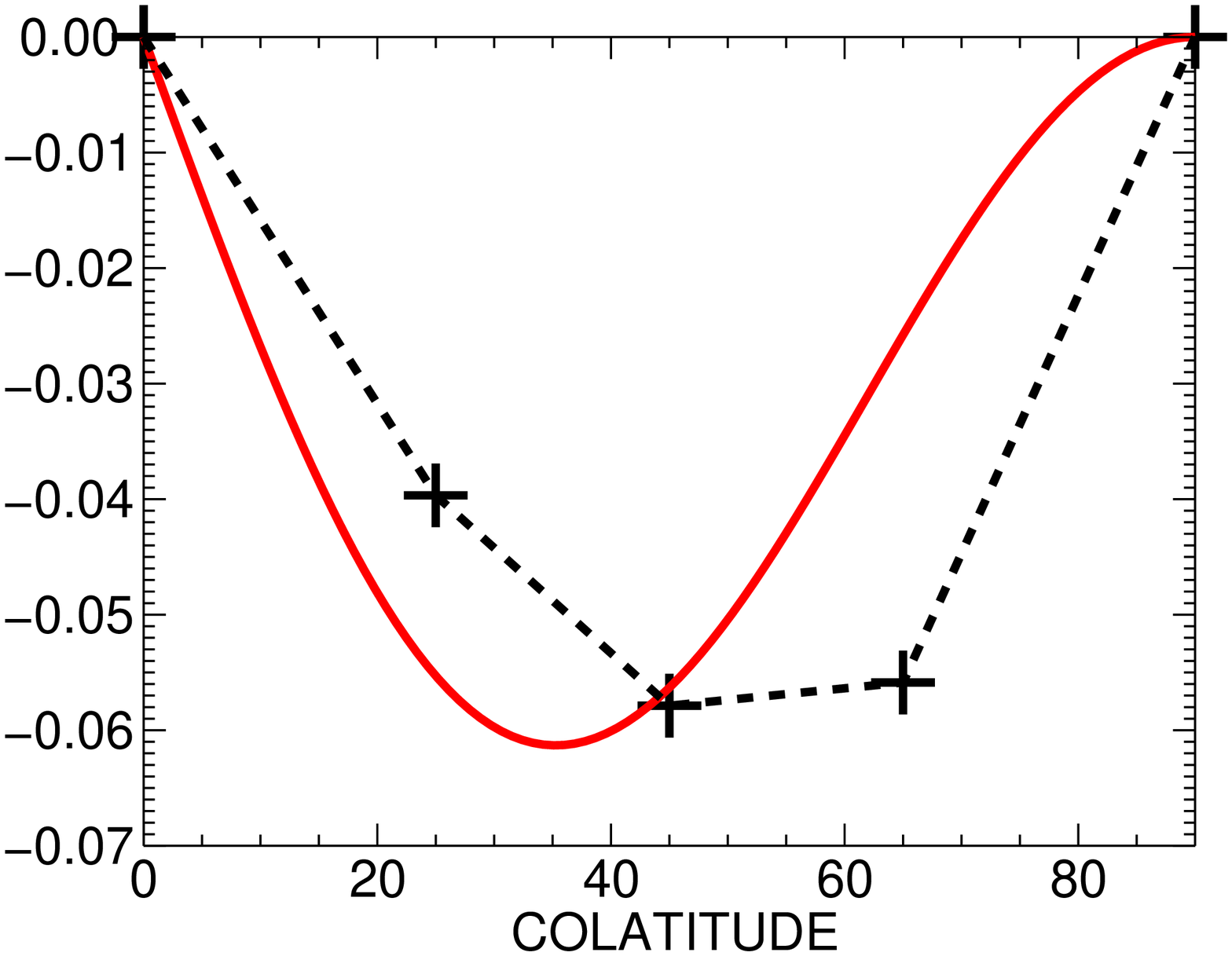}
\includegraphics[width=4.5cm]{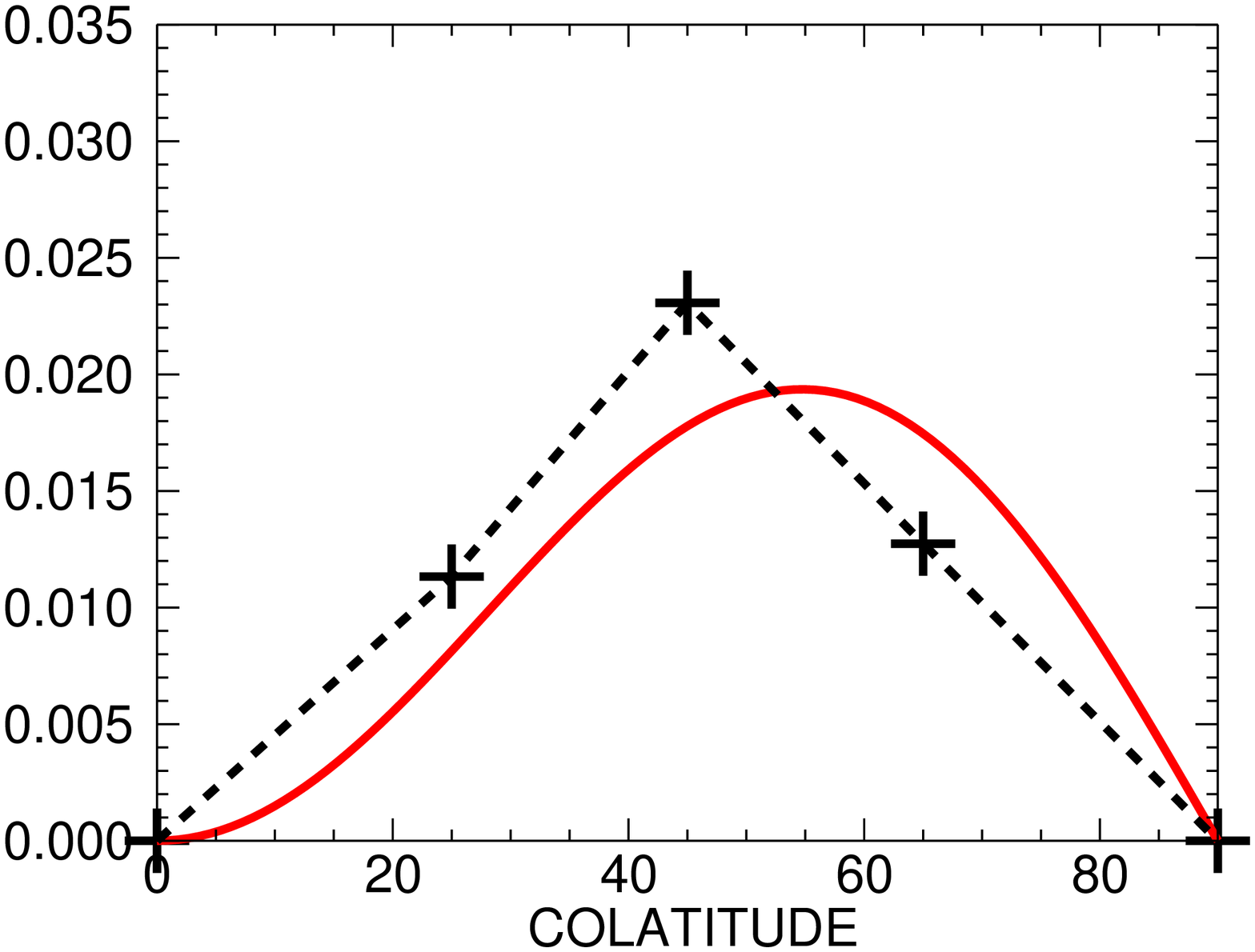}
}
\caption{Simulation results for the vertical angular momentum transport $T_{xz}$ (left) and the latitudinal  angular momentum transport $T_{yz}$ (right) without magnetic fields ($\vec B=0$) vs. colatitude $\theta$. The red lines give the $\theta$ functions used in Eqs. (\ref{VV}) and (\ref{HH}). The crosses at $\theta=45^\circ$  reflect the values of Table \ref{tab1} (first line); $\Om=10$,
 $\Prr=0.1$, and $\Pm=0.1$. }
\label{lambda}
\end{figure} 
Under the influence of magnetic fields, the values  are reduced (`magnetic quenching')  depending on the inclination of the background fields. We shall assume   in Eqs. (\ref{VV}) and (\ref{HH}) that the typical profiles  in   Fig. \ref{lambda} are not changed by the magnetic influences.

In order to formulate the magnetic influences on the angular momentum fluxes, we shall restrict { our computations}
to mid-latitudes ($\theta=45^\circ$).
\begin{figure*}[hbt]
 \vbox{
 \hbox{
 \includegraphics[width=4.45cm]{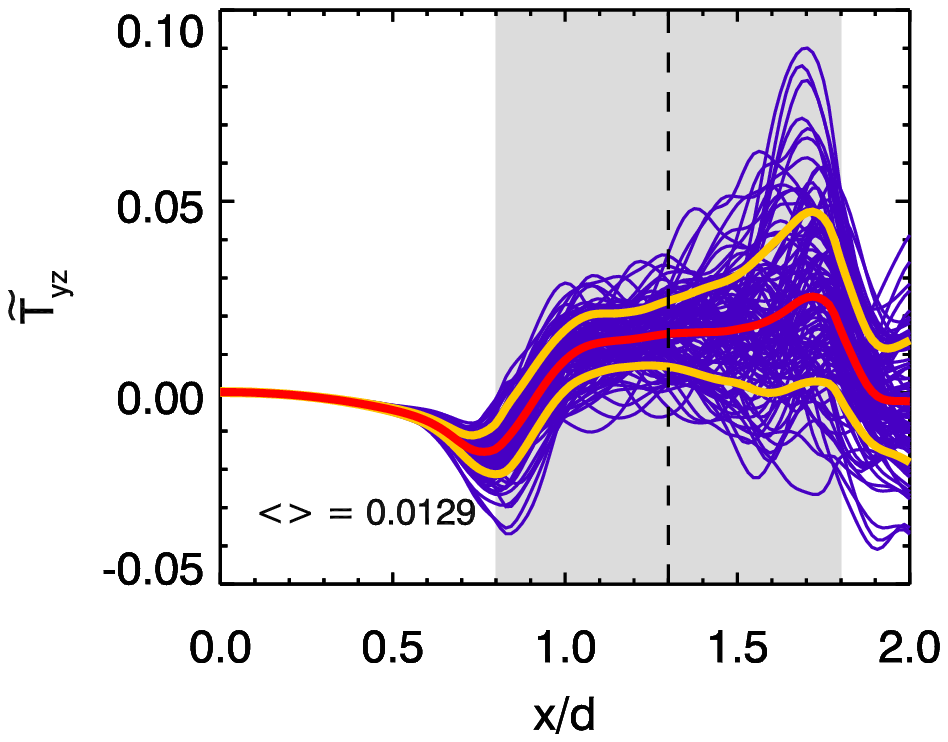}
\includegraphics[width=4.45cm]{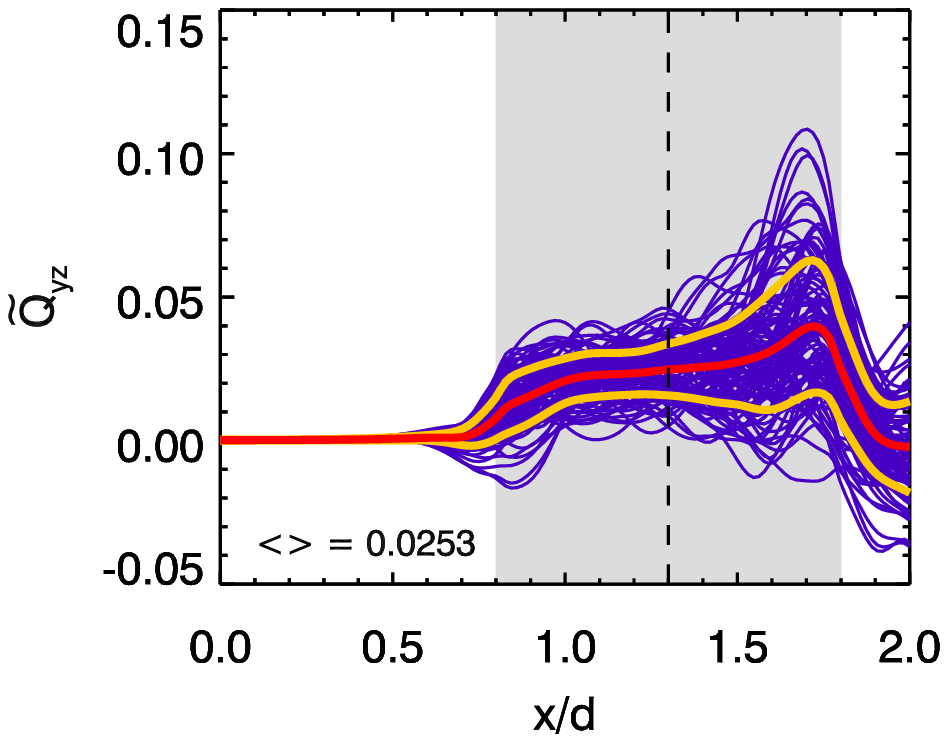}
\includegraphics[width=4.45cm]{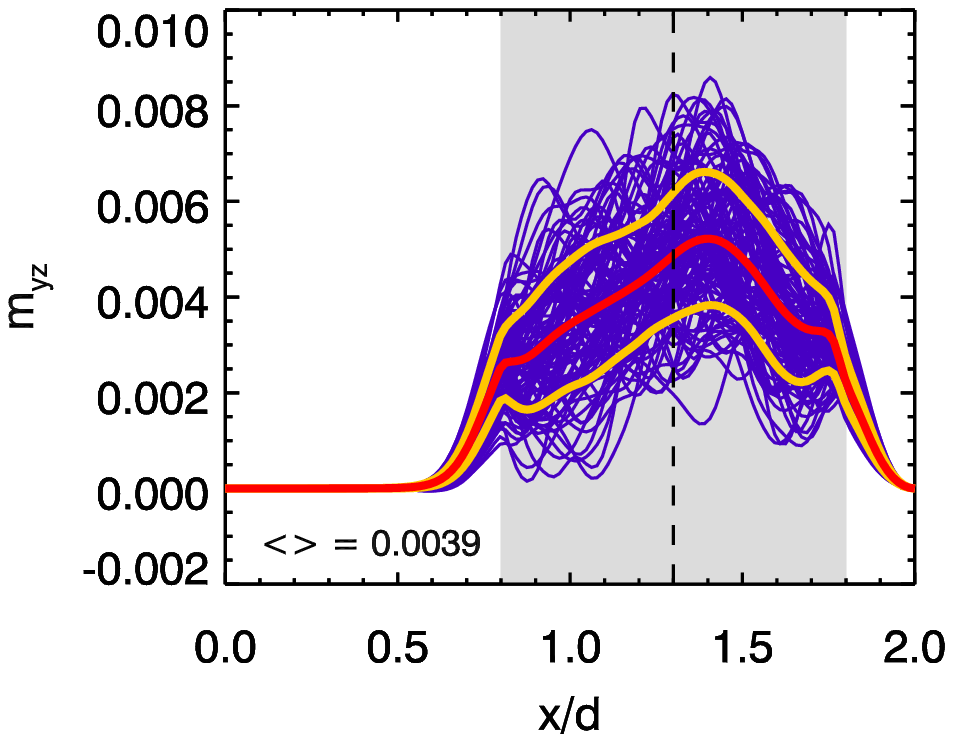}
\includegraphics[width=4.45cm]{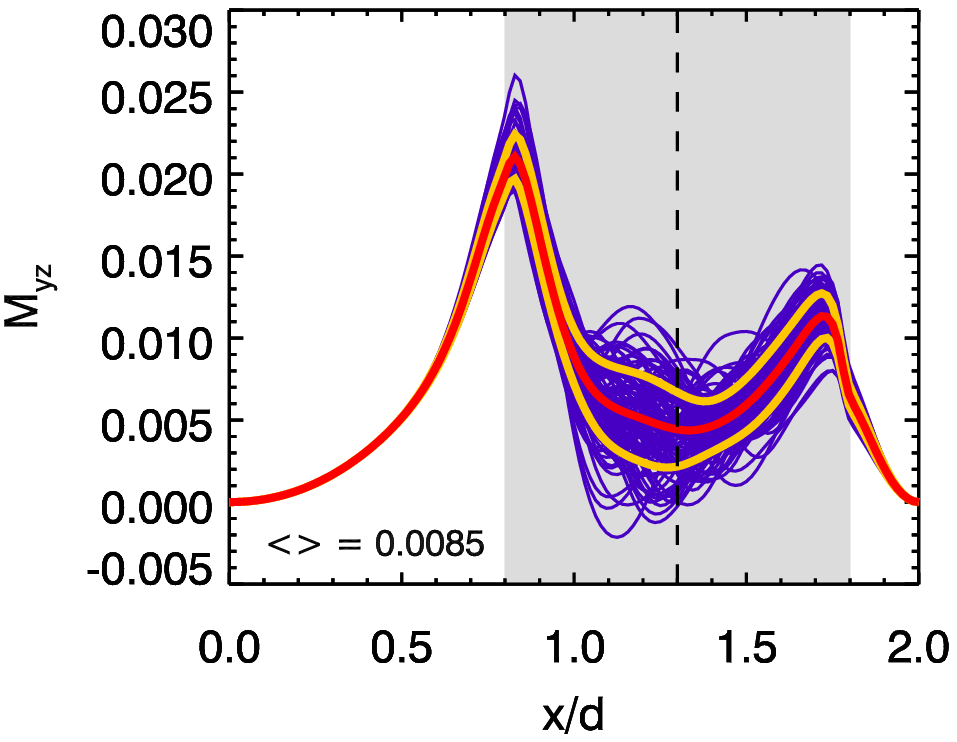}
}
\hbox{
 \includegraphics[width=4.45cm]{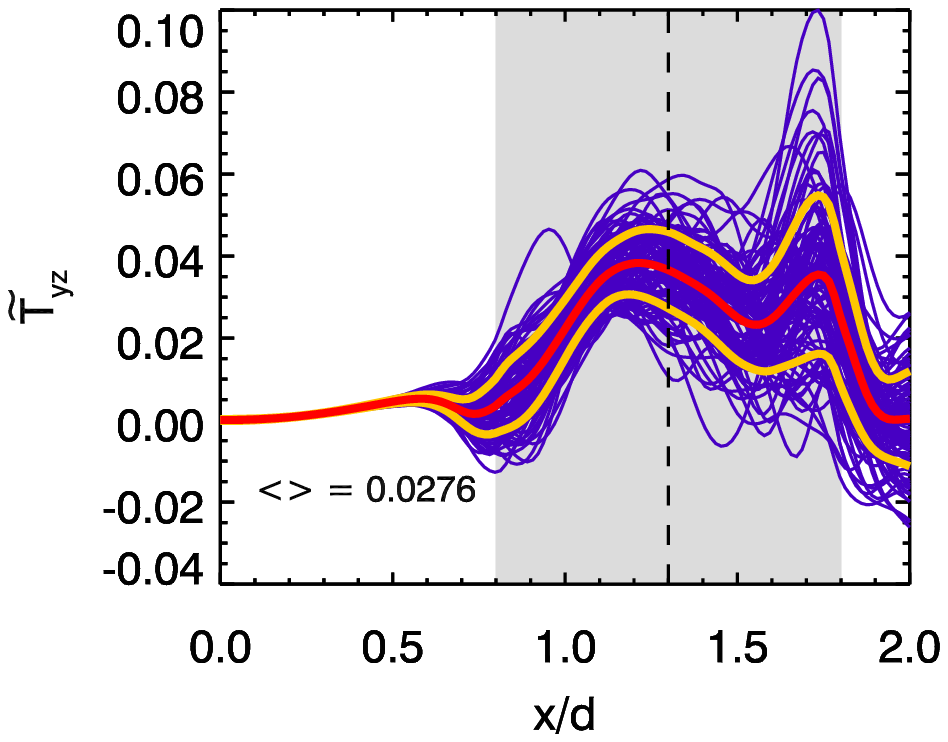}
\includegraphics[width=4.45cm]{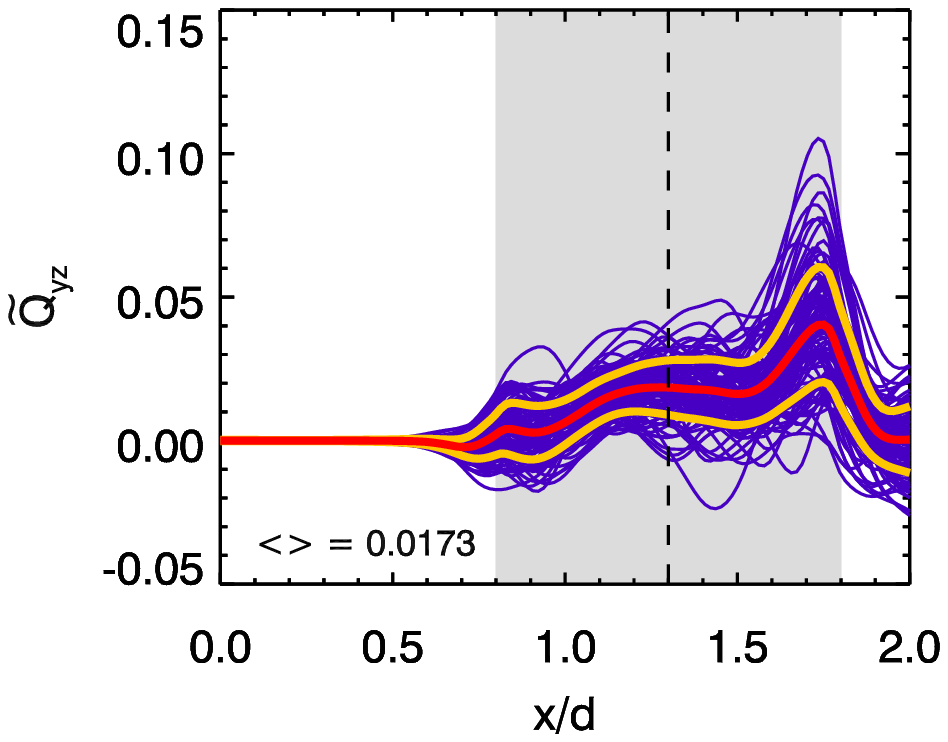}
\includegraphics[width=4.45cm]{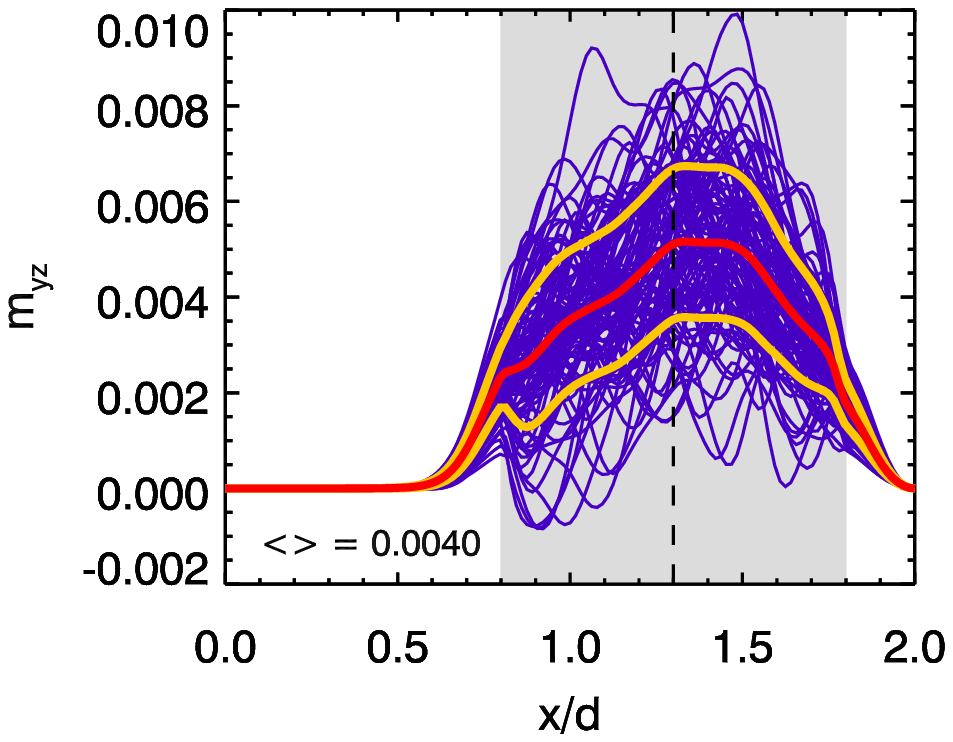}
\includegraphics[width=4.45cm]{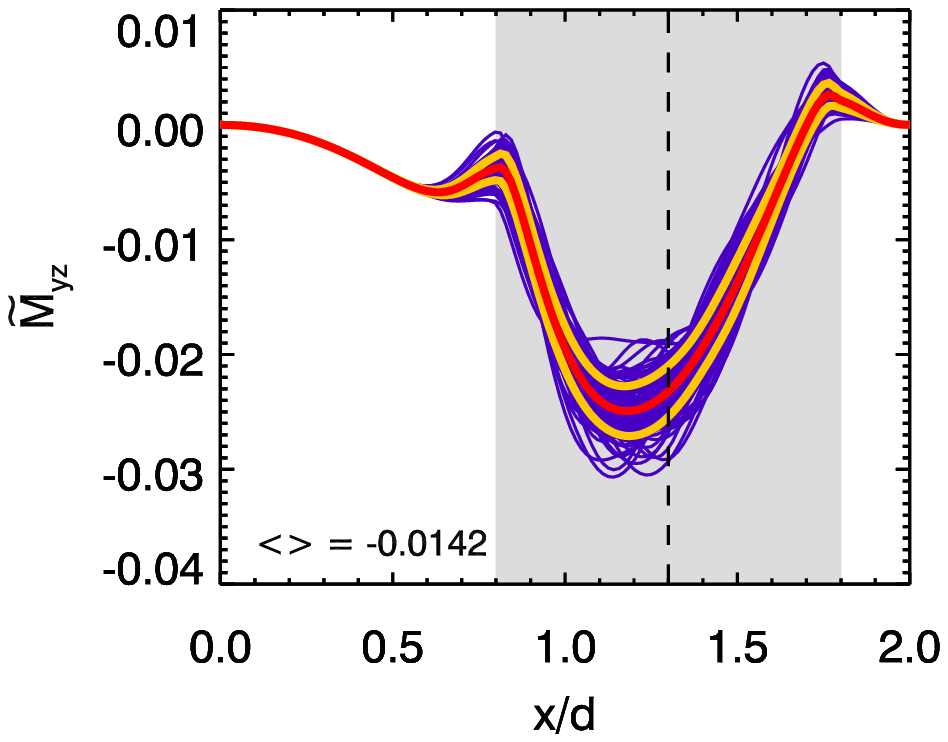}
}
}
\caption{Snapshots of the ($yz$)-correlations (normalised with $u_{\rm rms}^2=28$) for $\Om=10$ and  a magnetic field $\vec{B}=(0,1,10)$ (top) or $\vec{B}=(0,1,-10)$ (bottom). From left to right:  $\tilde T_{yz}, \tilde Q_{yz},\tilde  m_{yz}$, and $M_{yz}$. 
In opposition to nonrotating turbulence (see Fig. \ref{Om0hor}),   $T_{yz}$ is now positive for both field geometries. As \cite{KK17} show with their numerical dynamo simulations, the numerical values of the two middle rows are rather robust against the variation of the magnetic Prandtl number;  $\Prr=0.1$, $\Pm=0.1$, and $\theta=45^\circ$.}
\label{fig110}
\end{figure*} 
The  influences  on  the latitudinal angular momentum transport $T_{yz}$ are 
represented by Fig. \ref{fig110}, { providing the total stresses as well as the Reynolds and both  Maxwell stresses} 
for the two inclined  fields $\vec{B}=(0,\pm 1,10)$. In both cases, the small-scale Maxwell stresses are only weak and  the large-scale Maxwell stresses are of opposite sign. The main result is that for both magnetic configurations, the $T_{yz}$ are positive in opposition to  nonrotating turbulence (see Fig. \ref{Om0hor}),  hence with rotation   the Reynolds stresses dominate the Lorentz force. 

For $B_yB_z>0$, the Reynolds stress  grows and  for  $B_yB_z<0$ it sinks, that is the coefficient $D$ in Eq. (\ref{max1a}) is positive in opposition to $D^*$ in Eq. (\ref{max4}) which becomes negative. The  small numerical value  indicates a reduced effectivity of  the mean field  Lorentz force to transport angular momentum. It is  thus very  questionable that a  theoretical explanation of the solar torsional oscillation with only  the action of the dynamo-induced Lorentz force may work.

\begin{table}
\caption{
Results for the vertical flux  $\tilde T_{xz}$  (averaged over  radius and time) with $B_z=10$, $\Om=10$, $\Prr=\Pm=0.1$, and $\theta=45^\circ$.
}
\label{tab3}
\centering
\begin{tabular}{l||c|ccccc}
\hline\hline

\\

$B_z $& $B_x$& $\tilde T_{xz}$ & $\tilde{T}^*_{xz}$& $\tilde D^*$&  $T_{xz}/Q_{xz}|_{B=0}$& $\tilde T_{xz}|_{\Om= 0}$\\ \\
\hline
&&&&\\
0&0 &  &-0.062&  & & \\
\\
10&1 & -0.059 &  &&{ 0.95}&0.0023 \\ 
&& &-0.049 & -0.013   & &\\ 
10&-1&-0.038 & &   &0.61&-0.0023 \\ \\
 10&10 &-0.15 &  & &    2.4 &0.023 \\ 
  &&  & -0.029 &-0.014&   &  \\ 
 10&-10& 0.09 &  &&      -1.4 & -0.023\\ \\
  \hline
\end{tabular}
\end{table}
 
The code  was also probed with the radial flux $T_{xz}$ for  an inclined   magnetic  field $B_z=10$ 
with components $B_x$ varying between  $\pm 1$ and $\pm 10$ (Table \ref{tab3}). Here the nonmagnetic value of $\tilde{T}^*_{xz}= -0.062$ was magnetically reduced to $\tilde{T}^*_{xz}=-0.029$ by the field $\vec{B}=(\pm 10,0,10)$. 
This is a clear but  mild magnetic quenching. 
For $B_x=1$ and $B_z=10$, the equipartion value of the field is $\beta=0.50$ and for   $B_x=B_z=10$ it  is $\beta=0.75$ with $\beta^2$ as the ratio of magnetic to  kinetic energy. 
The reported   quenching behaviour is   similar to that of \cite{K19} in a quite different approach. We also find that the coefficient $\tilde D^*$ is basically uninfluenced by the magnetic background field. For very strong fields the mean value $\tilde{T}^*_{xz}$ vanishes, while $\tilde D^*\to -0.018 $ known from Section \ref{No} for $\Om=0$. The main result is thus that for small fields $|B_x|$, the flux  $T_{xz}$ is always negative but for large $|B_x|$, when the Lorentz force dominates,  both signs appear indicating that the Reynolds stress is smaller than  the Maxwell stress.

All correlations are written in units of the reference turbulence intensity $u_{\rm rms}^2=28$ resulting  from the nonrotating and unmagnetised turbulent fluid. { With $\Om=10$, we  used a rather fast rotation rate in order to minimise the large-scale shear flows which develop in the turbulent box \citep{BHT98}. The faster the rotation is, the weaker the shear 
of these flows and the smaller their influence  onto the calculated correlations. As demonstrated by \cite{RKK19}, both the calculated nondiffusive fluxes $V$ and $H$ are  minimum} values; they would  be slightly  larger under suppression  of the action of the induced shear flows. 

We note that  the  modified Lorentz force factor $D^*$ is   always negative  and 
 (almost) independent of the magnetic  inclination angle. It only grows for very small  angles,
 and  it depends  on $B_x$ only for $|B_x|\ll |B_z|$ when its amount becomes smaller. The dependence on $B_x$  vanishes for larger values. The consequence is that for  $|B_x|\ll |B_z|$ the magnetic quenching of $Q_{xz}$ cannot be compensated for by the Lorentz force $B_xB_z$ so that the total sum $T_{xz}$ becomes   smaller than the unmagnetised value $Q_{xz}(B=0)$ as shown in the next to last column of Table \ref{tab3}. This behaviour  is the natural counterpart to the phenomenon of  negative effective-magnetic-pressure where  total pressure is {also  reduced} by the magnetic parts, but only for a weak  magnetic field \citep{BK11,KBK12,RKS12}. Here, the magnetic-induced  reduction of the total  angular momentum transport  happens for  $B_x\lsim 0.1 B_z$ (similar for $B_y$).

The main results of the simulations for  $B_z=20$ and various $B_x$ and $B_y$ are summarised in  Table \ref{tab1}. The reference values of  the nonmagnetic approximation   are  given in the first row of the table.  
The last columns in the table (also in boldface) give the fluxes due to  the  Lorentz force alone without rotation.  The underlined numbers 
provide the resulting angular momentum fluxes  with rotation and the equipartition  field  $B_z=20$ and $B_x=\pm 0.1$ or   $B_y=\pm 0.1$.  For the given rotation rate and the magnetic field geometry,  the resulting angular momentum transport is dominated by the turbulence rather than by the Lorentz force of the background field. 
The magnetic influence on the total stresses  is, nevertheless, remarkably strong. The interplay of rotation, magnetic fields, and turbulence  mainly leads to a suppression of the turbulence-induced angular momentum transport despite the contribution of the large-scale Lorentz force. The underlined numbers in Table \ref{tab1} demonstrate
how the turbulent transport  is suppressed by the inclined field with the given components.  The vertical transport $T_{xz}$ for both signs of $B_x$
lies below the hydrodynamic value. The horizontal transport $T_{yz}$ is also suppressed, but there is a  difference for positive and negative $B_y$.  In the latter case, the magnetic and the nonmagnetic $T_{yz}$ hardly differ, while for positive $B_y$ the suppression is very strong. The signs of the total fluxes never differ for different signs of the (small) inclination angles $p$.

\begin{table*}
\caption{Normalised angular momentum fluxes  $\tilde T_{xz}$  and  $\tilde T_{yz}$ (averaged over radius and time) with $B_z=20$ and for various meridional magnetic field components $B_x$ and $B_y$. 
The boldfaced numbers give the correlations without a field (first line) and without rotation (last columns); $\Om=10$, $\Prr=\Pm=0.1$, and $\theta=45^\circ$.
}
\label{tab1}
\centering
\begin{tabular}{l||ccccc||cccccc}
\hline\hline

\\

$B_z $& $B_x$&$p_x$& $\tilde T_{xz}$ & $\tilde{T}^*_{xz}$&    $\tilde T_{xz}|_{\Om=0}$ & $B_y$&$p_y$& 
$\tilde T_{yz}$& $\tilde{T}^*_{yz}$& $\tilde T_{yz}|_{\Om=0}$\\ \\
\hline
&&&&\\
0&0& -  &{ -0.0623}& - &0&  0 &-& {  0.0235}&-&0  \\
\\
20&0.05&0.0025 &-0.0541 &     & 0.00023&0.05 &0.0025  &0.0201  &&-0.0015\\ 
&&& &-0.051   & & && &0.023& \\ 
20&-0.05&-0.0025& -0.0475&    &-0.00023 &-0.05&-0.0025  &0.0262 & &0.0015\\ \\
20&0.1 &0.005&\underline {-0.0537} &  &{ 0.00046} & 0.1&0.005   & {\underline{ 0.0159}}    &&{ -0.0030}\\ 
&&& &-0.053 &   & & & &0.021&\\ 
20&-0.1&-0.005&{\underline{ -0.0498}} & &  { -0.00046} &-0.1&-0.005   &{\underline{ 0.0255}}    & &{ 0.0030}\\ \\
20&1 &0.05& -0.073 &  & 0.0046& 1 & 0.05&0.00829 &&-0.030 \\ 
&&& &-0.049 &  & &&  &0.025&\\ 
20&-1&-0.05&-0.026 &    &-0.0046 &-1 &-0.05 &0.0424& &0.030\\ \\
  \hline
  \end{tabular}
\end{table*}


\section{Rotation laws}
%
To obtain the (axisymmetric) rotation law $\Om=\Om(r,\theta)$ in the solar convection zone, 
 the equation for the conservation of angular momentum in spherical coordinates 
\beg
\rho r^2 \sin^2 \theta \frac{\partial \Om}{\partial t} + \div\ \vec{t}=0
\label{angmom}
\ende
was solved using a time-dependent finite difference code and the density profile of the solar convection zone with variations of three orders of magnitudes.
The relevant components of the angular momentum flux vector $\vec{t}$ are\begin{eqnarray}
  t_r   &   = &r \sin \theta \rho \left(Q^\nu_{r\phi} +T_{r \phi}\right), \\ 
  t_\theta & =& r \sin \theta \rho \left(Q^\nu_{\theta\phi} +T_{\theta \phi} \right)
  \label{angmom2}
  \end{eqnarray}
  with the viscous part of the stress tensor
\begin{eqnarray}
  Q^\nu_{r \phi}  =  -\nuT r \sin \theta  \frac{\partial \Om}{\partial r},  \ \ \ \ \ \ \ \ \ \ \ \ \ \ 
  Q^\nu_{\theta \phi}  = - \nuT \sin \theta  \frac{\partial \Om}{\partial \theta}
  \end{eqnarray}
 and constant viscosity coefficient $\nuT$. The boundary conditions at the top and bottom of the convection zone are simply $t_{r}=0$. They ensure that no  angular momentum leaves  the convection zone. We note that 
 the angular momentum transport by the  meridional flow was neglected here  so that only the azimuthal component (\ref{angmom}) of the Reynolds equation must have been solved. The complete mean-field theory on the basis of Eq.\ (\ref{angmom2}) provides one-cell meridional flows   circulating counterclockwise in the northern hemisphere \citep{RK13} close to the recent results of helioseismology \citep{G20}.  
 
 The code started with rigid rotation and was run until a stationary solution was reached.
The latitudinal profiles  in Fig. \ref{lambda} are the basis for the formulation
\beg
{\tilde T}_{xz} =  V \cos^2\theta \sin\theta
\label{VV}
\ende
with the minimum at $\theta=35^\circ$ and 
\beg 
{\tilde T}_{yz}= H \cos\theta \sin^2\theta,
\label{HH}
\ende
with the maximum at $\theta=54^\circ$. The $V$ and $H$ are dimensionless quantities. They are coupled 
to the viscosity term in the angular momentum equation by a dimensionless factor $\zeta=u^2_{\rm rms}/\nu_{\rm T}\Om$ for which solar values provide $\zeta\lsim 10$.
The parameter $\zeta$  includes all normalisation factors, hence $T_{r\phi}:=\zeta \tilde T_{xz}$ and  $T_{\theta\phi}:=\zeta \tilde T_{yz}$.

Equation (\ref{VV}) contains the nontrivial assumption  that the radial turbulent transport vanishes at the equator, which is the case for quasilinear analytical calculations without magnetic fields \citep{RK13}.  We do not expect contradictions by the magnetic fields as the toroidal background field vanishes at the equator. However, without meridional circulation and without turbulence-induced anisotropic heat transport, the immediate consequence of vanishing radial flux along $\theta=90^\circ$ is the uniformity of $\Om$ in the equatorial plane in contrast to the slight superrotation which is  observed with methods of helioseismology. This little disadvantage of our model may be accepted by the reader.

The  coefficients $V$ and $H$ in Eqs. (\ref{VV}) and (\ref{HH})   were calculated from  the numerical results in Table \ref{tab1}  for $\theta=45^\circ$.
\begin{figure}[ht]
  \hbox{
  \includegraphics[width=4.5cm]{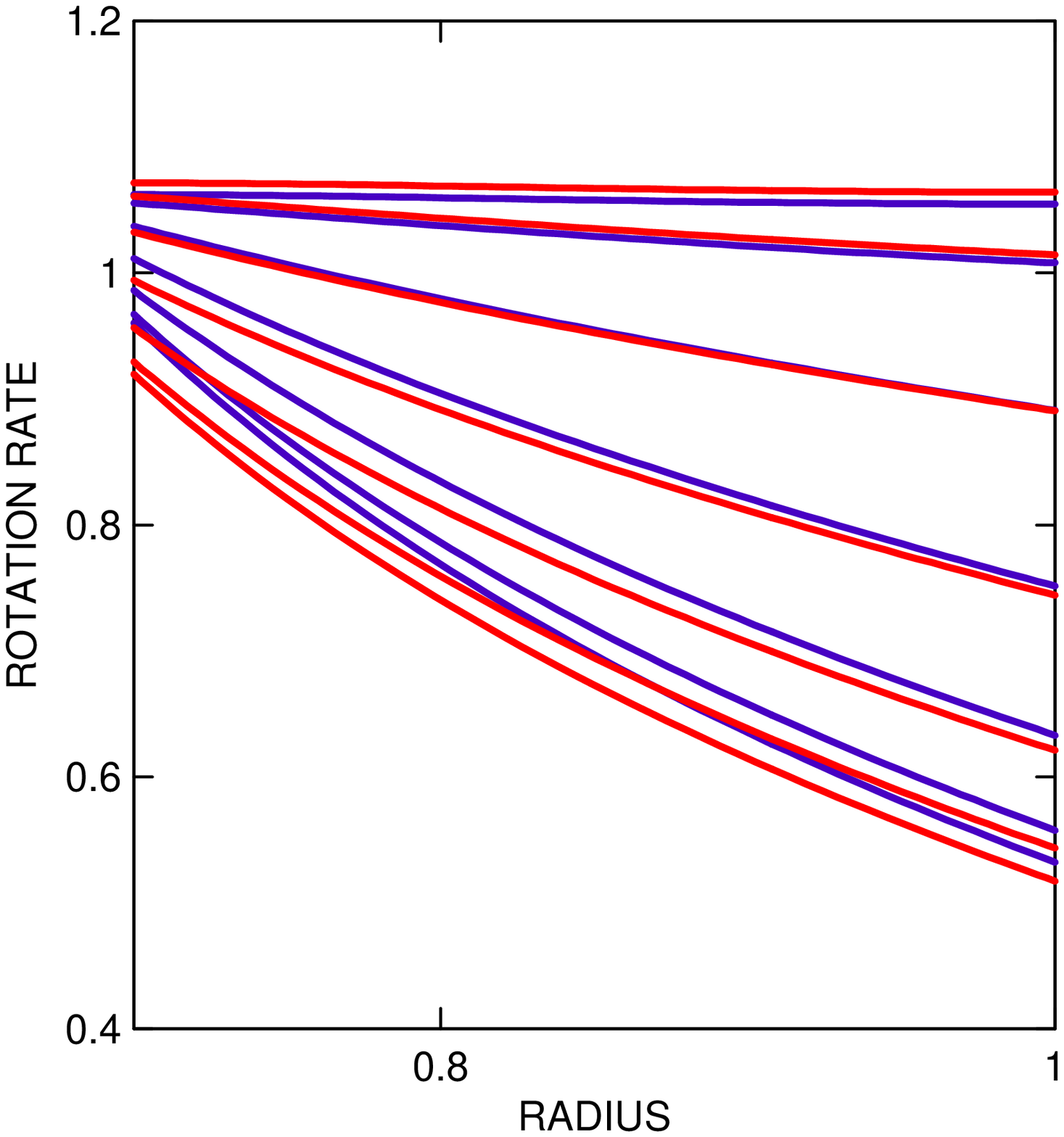}
 \includegraphics[width=4.5cm]{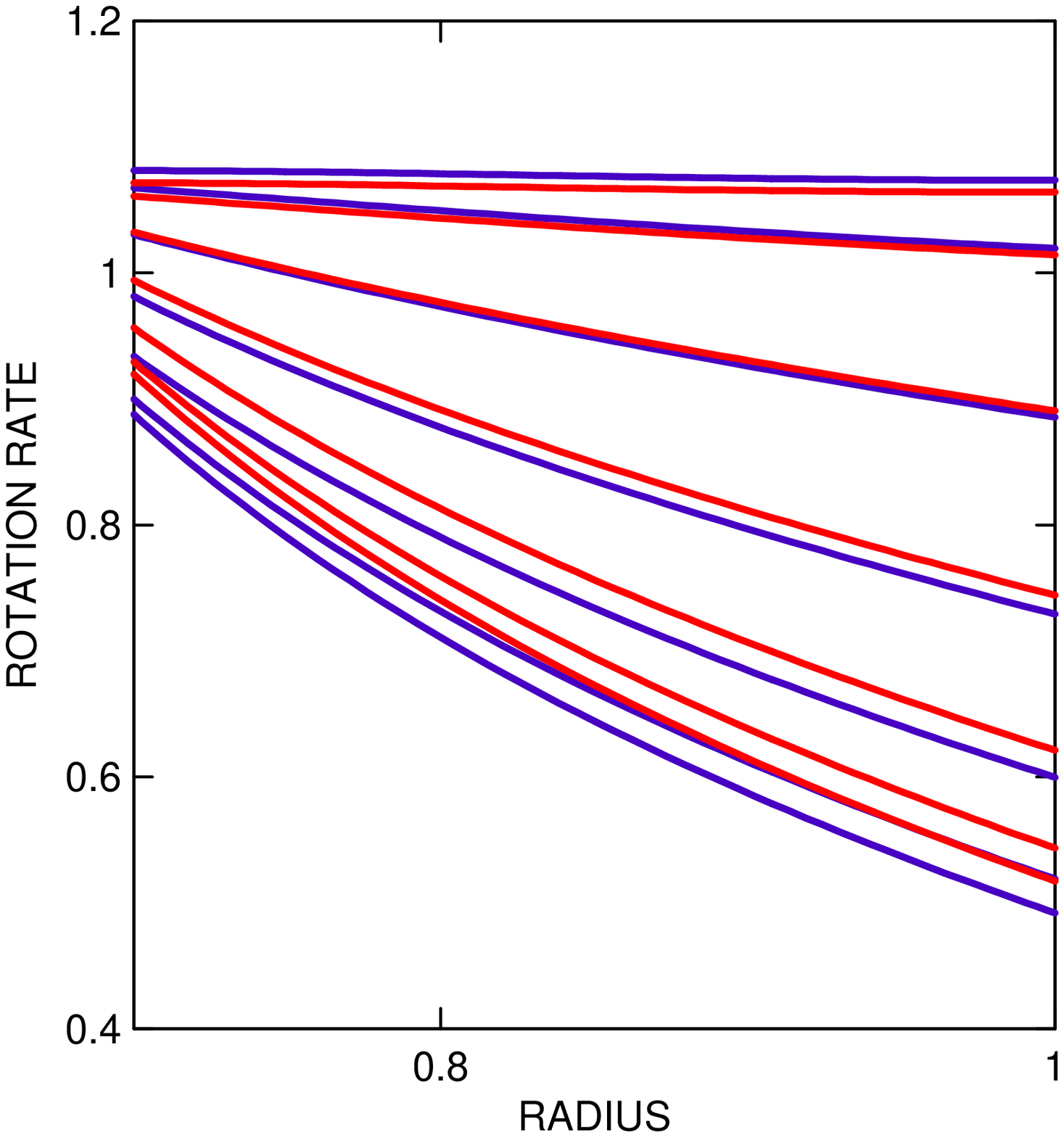}}
\caption{Rotation laws  in radius $r$ and colatitude $\theta$ for fast rotation ($\Om=10$) under the influence of the large-scale Lorentz force  for  $\vec{B}=(0,0.1,20)$ (left panel, blue lines) and $\vec{B}=(0,-0.1,20)$ (right panel, blue lines). The differential rotation  is due to the nonmagnetic $\Lambda$ effect (red lines). The uppermost $\Om$-isolines belong to the equator, while the lowest curves belong to the poles.}\label{rotlaw0}
\end{figure} 
For a rotating but  nonmagnetic box, one finds from  the first line of Table \ref{tab1}  the values  $V=-0.18$ and $H=0.066$. The radial flux exceeds the latitudinal one by a factor of three. 
The resulting isolines of the angular velocity $\Om(r,\theta)$ are given by a red colour   in the plots of   Fig. \ref{rotlaw0}. In the spirit of the Malkus-Proctor approximation,   the nonmagnetic  rotation law may be  modified only by the large-scale Lorentz force of the two  fields  $\vec{B}=(0,\pm 0.1,20)$ (blue lines). 
We note again that $B_z=20$  represents the equipartition value of the azimuthal field, while $B_y=\pm 0.1$ is still a rather large  value for the possible latitudinal field component. 
The formally resulting momentum fluxes  are $H=0.058$
for $B_y=0.1$ and $H=0.075$  for $B_y=-0.1$. Fields with a positive inclination angle transport the angular momentum poleward and  fields with a negative inclination angle transport the angular momentum equatorward. In general,  the field with the positive inclination angle (as in the solar convection zone) reduces the latitudinal shear, while the field with the negative inclination angle enhances the shear.  There is even a tendency of a  magnetic deceleration of the equatorial rotation rate but this is only a very  small effect which becomes even  weaker as $B_y$ diminishes.
 To include the  magnetic background fields in a consistent way,  the underlined coefficients in Table \ref{tab1} must be used  to evaluate  the coefficients $V$ and $H$ for $ p_x=p_y=\pm 5 \cdot 10^{-3}$. 
 In  Fig. \ref{rotlaw1}
 the blue lines again represent the rotation laws under the influence of these magnetic fields, while the red lines  are valid for $\vec B=0$.  
  
\begin{figure}[ht]
\vbox{
 \hbox{
  \includegraphics[width=4.5cm]{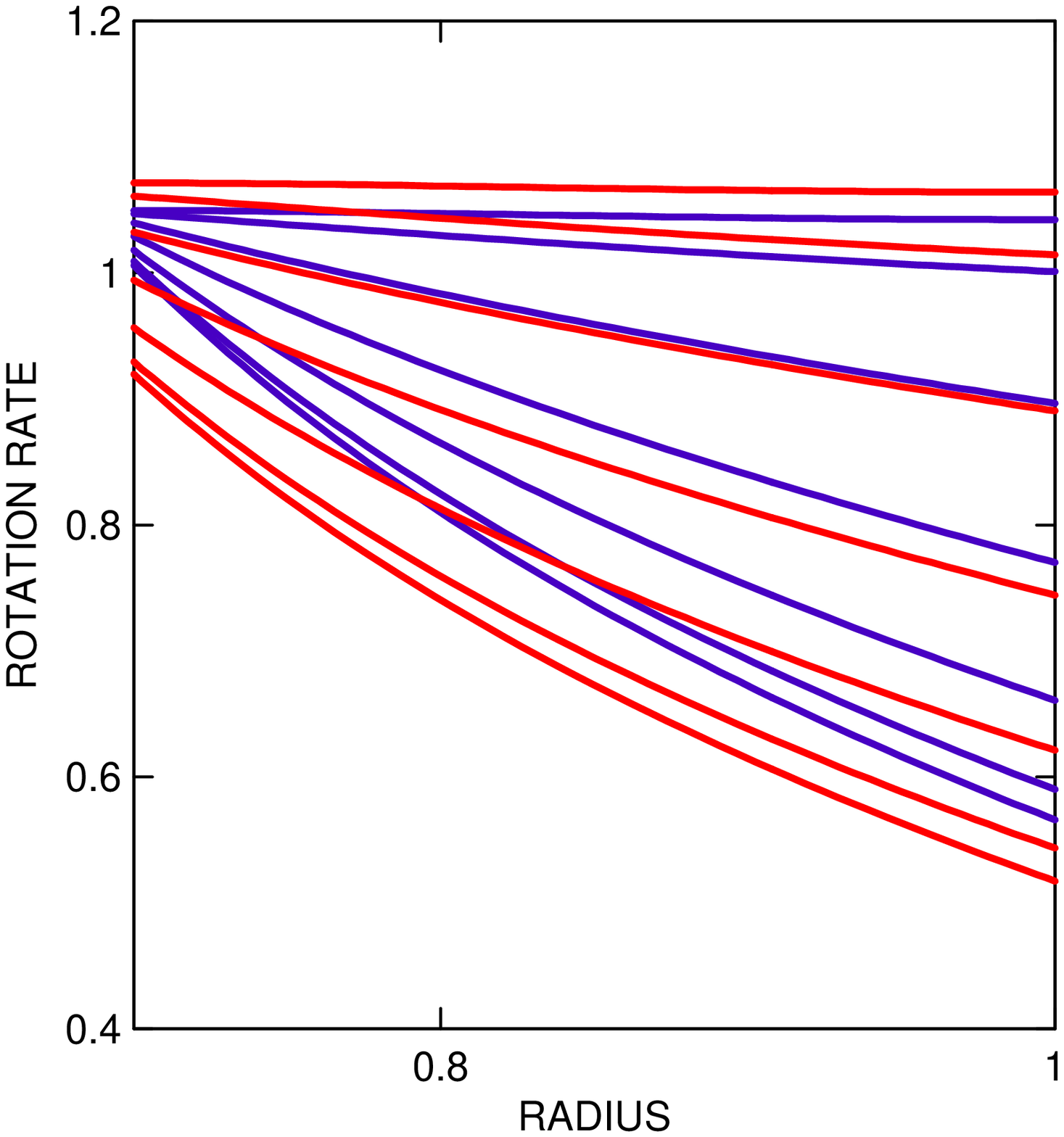}
 \includegraphics[width=4.5cm]{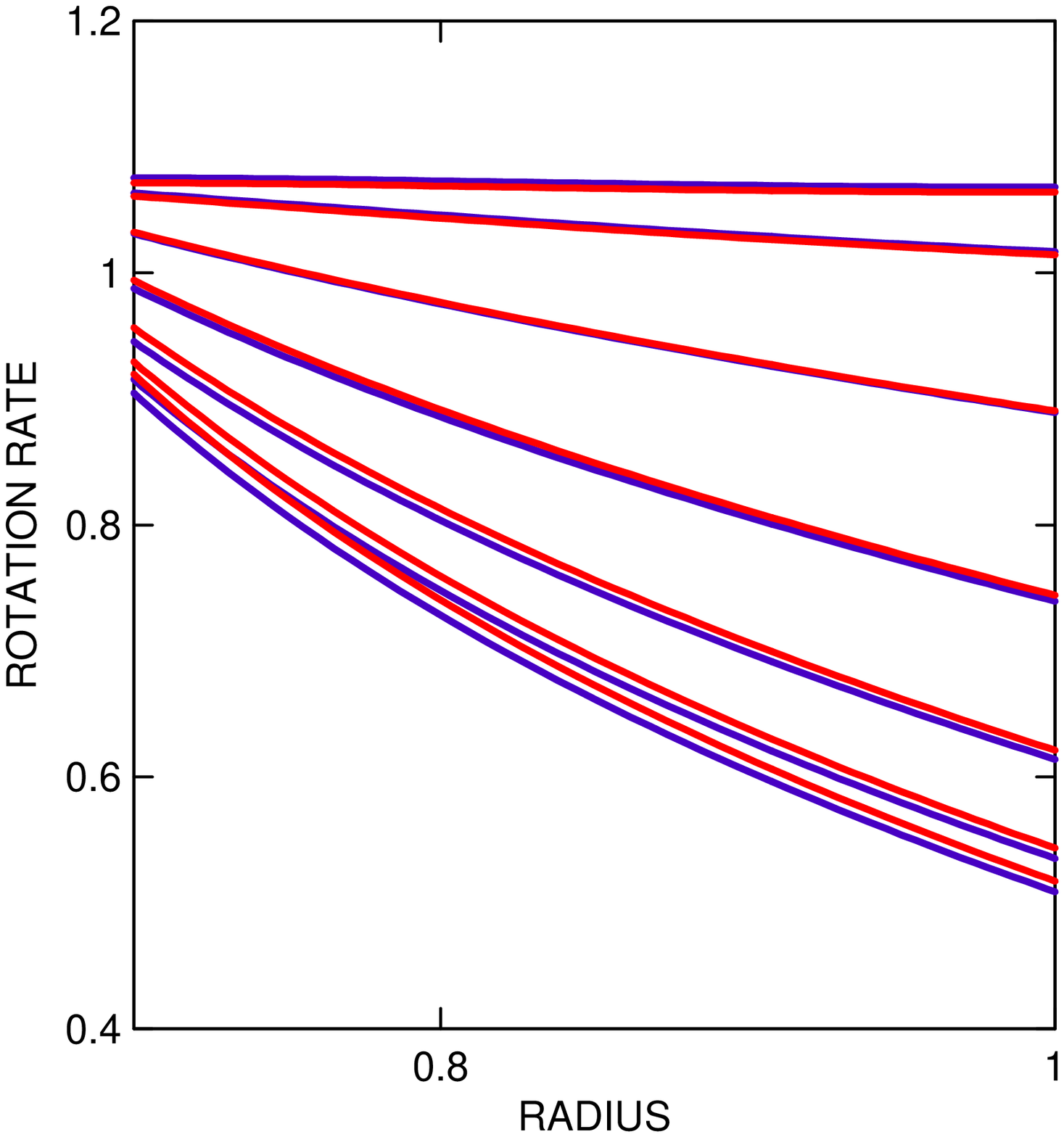}} 
 \hbox{
  \includegraphics[width=4.5cm]{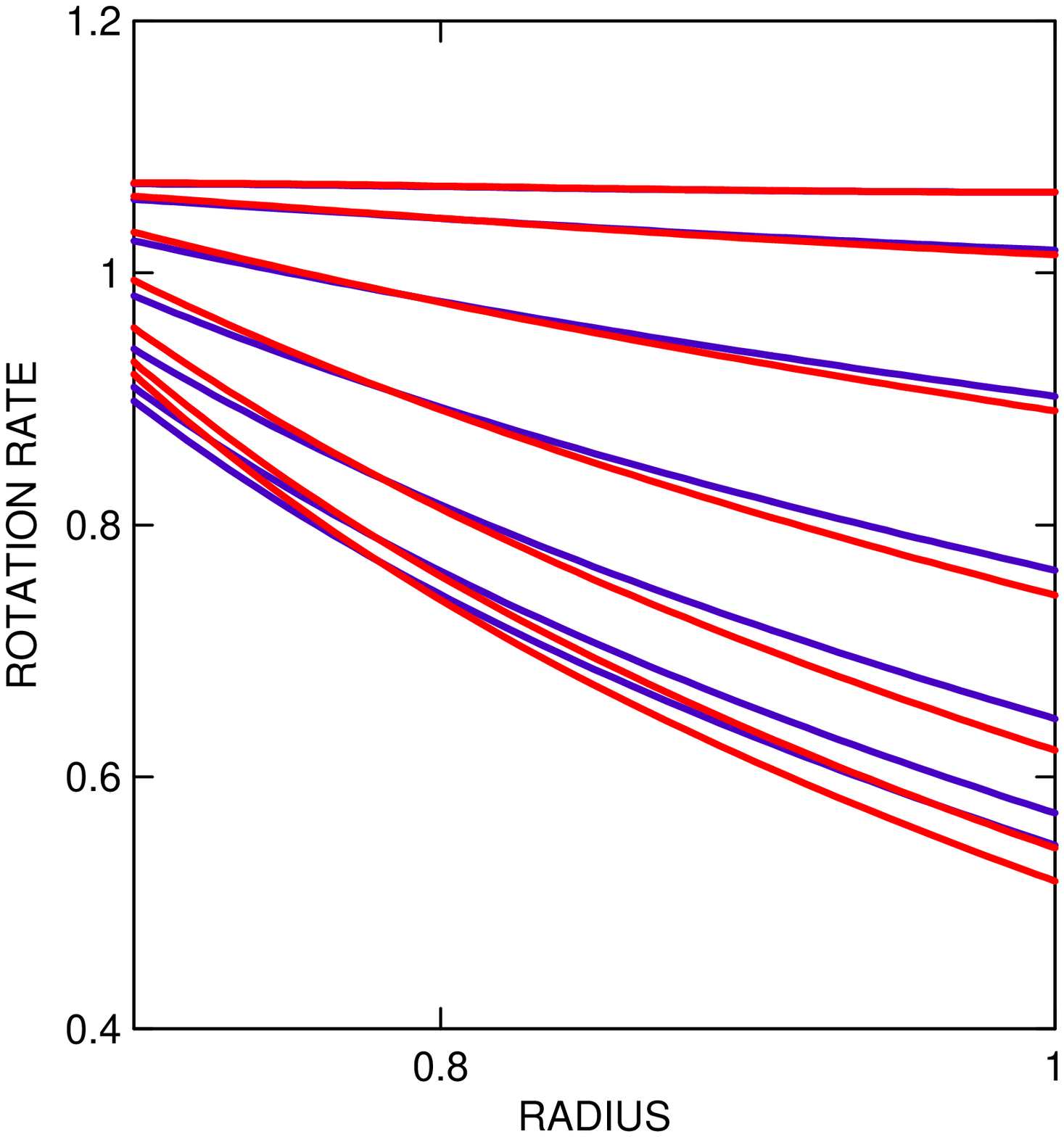}
 \includegraphics[width=4.5cm]{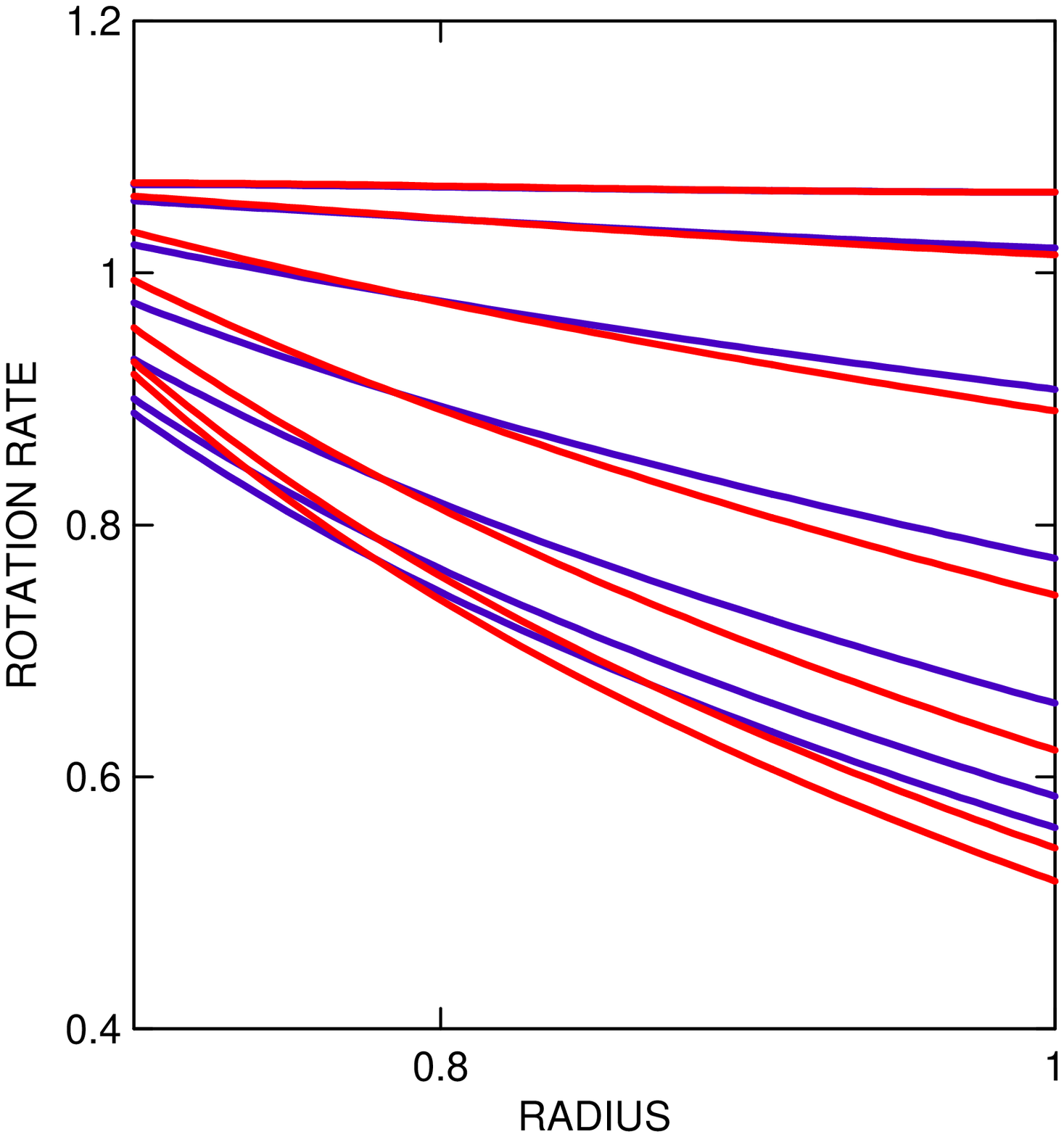}} }
\caption{Similar to Fig. \ref{rotlaw0}, but with the total magnetic stress-tensor from Eq. (\ref{3c}). The applied magnetic fields (at $\theta=45^\circ$) are  $\vec{B}=(0,0.1,20)$ (top left, model 1), $\vec{B}=(0,-0.1,20)$ (top right, model 2),  $\vec{B}=(0.1,0,20)$ (bottom left, model 3), and $\vec{B}=(-0.1,0,20)$ (bottom right, model 4). The red  curves are the nonmagnetic rotation laws.}
\label{rotlaw1}
\end{figure} 

In Table \ref{tab4}, the four models with different magnetic  geometries  are summarised for which the rotation laws have been calculated. All models work with the equipartition value $B_z=20$ and with the inclination angles  $|p|=5\cdot 10^{-3}$. We focus our attention  to the model 1 with positive $B_y B_z$,  which  should be most reliable  for the solar convection zone. The   combinations of $B_x$ and $B_z$ are much more speculative.

\begin{table}
\caption{Vertical and the horizontal fluxes of angular momentum for 
various combinations of the meridional magnetic field components;
  $\Om=10$ and $\Prr=\Pm=0.1$.
}
\label{tab4}
\centering
\begin{tabular}{l|ccccc}
\hline\hline

\\

model & $B_x$ & $B_y$ &$B_z$& $V$&$H$\\ \\
\hline
&&&&\\
0 & 0 &0&0&{-0.18}& {0.066}  \\ \\
1 & 0 &{0.1}&20&{-0.18}& {0.045}  \\
2&0&-0.1&20&-0.18& 0.072 \\ \\
3 &0.1 & 0 &20& -0.15&0.066  \\
 4&- 0.1 &  0&20& -.0.14&0.066  \\ \\
  \hline
\end{tabular}
\end{table}

The  rotation laws for models 1--4 are displayed in Fig. \ref{rotlaw1}. The lines represent the rotation rates within the convection zone for various latitudes. The uppermost profiles belong to the equator, while the lowest lines belong to the poles. All models provide the magnetic-suppression of the equator-pole difference of the surface rotation.  The spread of the (non-magnetic) red lines  exceeds  that  of the (magnetic) blue lines for all models except for model 2 where the magnetic influence is basically  weak.  Only in one case, however, is the equatorial rotation rate also magnetically reduced.
This is true for model 1 where the upper blue line clearly  lies below the upper red line, hence the equator rotates slower during the magnetic activity maximum. For all other magnetic configurations, the equatorial rotation rate is magnetically uninfluenced. Obviously, the deceleration of the equator is a consequence of the magnetic  quenching which reduces the horizontal transport coefficient down  to $H=0.045$.
Model 1 shows   the magnetic-originated variation of the equator-pole difference of $\Om$ as 3--4 units of the magnetic-originated variation of the equatorial rotation rate so that  not only the first observational  condition { as shown in Eq.\ (\ref{Om2})}
  but also  the second one will  approximately  be fulfilled by  model 1.
%
%
\section{Results and discussion}
%
%
The explanation of a possible non-regular behaviour in the magnetic activity cycle has been started including the magnetic feedback on the internal solar rotation.
{ \cite{JW91}, \cite{KR94}, and \cite{T96} }
 added the conservation law of angular momentum
in the turbulent convection zone to the dynamo equations  including  magnetic feedback in order to simulate the interplay of induced magnetic fields and rotation. The form of the rotation law can be influenced by the magnetic field  in two ways.  If the magnetic field only consists on a (strong) toroidal field, then the Reynolds stress ($\Lambda$ effect and/or  eddy viscosity) is magnetically quenched. If, however, the magnetic field also possess (weak) latitudinal components then  small-scale as well as large-scale Maxwell stresses additionally contribute to the  angular momentum transport.
{ \cite{Y81}, \cite{S81}, \cite{NW84}, and \cite{RT86} }
presented the  first dynamo models   where large-scale Lorentz forces due to induced  magnetic fields modified the internal rotation laws. 

In the present paper, the total angular momentum transport by rotating  magnetoconvection under the influence of prescribed magnetic background fields was calculated by means of 3D MHD box simulations. They are motivated by the observational result that in the solar activity maximum both the equator-pole difference of the surface rotation rate and the equatorial rotation rate is reduced  where the latter has only a 1\% effect. The question is whether a combination of a (strong) azimuthal field component as well as (weak) radial and latitudinal field components can reproduce these observations.

A natural start of the calculations is  given  by Fig. \ref{rotlaw0}. The magnetic modifications of the rotation profiles are only due to the large-scale Lorentz force of two prescribed large-scale fields, which is the original configuration used by \cite{MP75}. The (dominating)
azimuthal field  with energy in  equipartition with the kinetic energy of the turbulence is $B_z=20$. If the extra latitudinal field $B_y=0.1$ is added, the resulting Maxwell stress transports the angular momentum poleward
and the latitudinal shear is magnetically reduced. It is clear that for $B_y=-0.1$, the opposite is true: The shear is  increased. These results are almost trivial and  for decreasing $|B_y|$ they disappear.

The calculations have been  repeated with the total stresses shown in Eqs. (\ref{Tr}) and (\ref{Tf}), which are  known from box simulations for rotating convection under the influence of   magnetic background  fields  of the same geometry. Then Reynolds stress, small-scale Maxwell stress, and large-scale Maxwell stress are the transporters of the angular momentum. The numerical  simulations  lead to the two main results that i) the influence of the  large-scale Maxwell stress is only small, that is 
 $T_{yz}\gg T_{yz}|_{\Om=0}$,
and ii) the magnetic influence on the Reynolds stress  depends on the sign of $B_y$, that is
$T_{yz}<T_{yz}|_{B=0}$ for $B_y=0.1$  and $T_{yz}\gsim T_{yz}|_{B=0}$ for $B_y=-0.1$. 
Consequently, the equator-pole difference of $\Om$  is almost not influenced if $B_y B_z<0$, but it is  strongly reduced if $B_y B_z>0$ (see top panel of Fig. \ref{rotlaw1}).  In the latter case, the equator is also decelerated in the activity maximum. Positive $B_yB_z$ is  expected  to exist within  the solar convection zone and the resulting magnetic behaviour of the rotation law with an accelerated equator fully complies with   the observations shown in Eq. (\ref{Om2}). 
The scenario appears to be consistent in the following case: If magnetic fields $B_\theta$ exist,  the turbulence-originated rotational shear produces positive products of $B_\theta B_\phi$ which, in agreement with the  observations, simultaneously reduce the   equatorial value of the  angular velocity and the surface differential rotation. It is also obvious that for the Sun the angular momentum transport by the Maxwell stress does not exceed the  transport by the Reynolds stress. With $u_r \simeq  u_\phi \simeq 100$~m/s and with $B_r\simeq 1$~Gauss and $B_\phi\simeq 10^4$~Gauss, one finds the ratio $B_r B_\phi/(\mu_0\rho c u_ru_\phi)\lsim$~O$(10^{-2})$ where  the correlation factor $c\simeq 0.1$ and the density $\rho\simeq 10   ^{-2}$~gcm$^{-3}$ have approximately been used. Indeed,  the solar observations reveal  the cycle-dependent
velocity variations as  never exceeding the $10^{-2}$ limit. One  would need toroidal fields of the order of $10^{5-6}$~Gauss in order to find the rotation law as basically modulated by the Lorentz force. 

As our model of the solar differential rotation is stationary and the latitude dependence of the Reynolds stress is completely determined through the $V$ and $H$ coefficients, it cannot be directly used for the explanation  of the torsional oscillation patterns. \cite{Howard1980} determined a global rotation law, 
which they subtracted from the observed profile of the rotation rate. The residual shows a pattern of alternating bands of fast and slow rotation. Zones of more rapid rotation appear at high latitude and migrate towards the equator similar to the active regions. More recent observations revealed  that the zone of faster than average rotation reaches the equator in the solar minimum  \citep{Basu2003,HH11}. This is in line with the  findings in the present paper.
\bibliographystyle{aa}
\bibliography{pressurekuek}

\end{document}